\documentclass[a4paper,11pt]{article}

\usepackage{graphicx}
\usepackage{amssymb}
\usepackage{float}
\usepackage{amsmath}
\usepackage{amsfonts}
\usepackage{dcolumn}
\usepackage{subfigure}
\usepackage{mathtools}
\pdfoutput=1 
\usepackage{jheppub} 
\usepackage[T1]{fontenc} 

\title{Critical collapse in K-essence models}

\author[a]{Radouane Gannouji}
\author[b, 1]{and Yolbeiker Rodr\'iguez Baez \note{Corresponding author.}}

\affiliation[a]{Instituto de F\'isica, Pontificia Universidad Cat\'olica de Valpara\'iso, \\Av. Brasil 2950, Valpara\'iso, Chile}
\affiliation[b]{Universidad T\'ecnica Federico Santa Mar\'ia, \\ Av. Espa\~na 1680, Valpara\'iso, Chile}

\emailAdd{radouane.gannouji@pucv.cl}
\emailAdd{yolbeiker.rodriguez@gmail.com}

\abstract{We study gravitational collapse in K-essence model with shift symmetry. For these models, we have the formation of two types of horizons, event and sonic. For the particular case $K(X)=X+\beta X^2$ we found three different regimes. In the weak field regime the scalar field disperses to infinity, in the very strong regime both horizons form at the same time and finally for the intermediate regime, the sonic horizon could form first or both horizons form at the same time. The threshold of formation of the horizon is found in the regime where the sonic horizon forms first. We observe a universal behavior with a scaling parameter $\gamma\simeq 0.51$. Interestingly this universal behavior is encoded in the sonic horizon even if an event horizon is expected to never form because of loss of hyperbolicity of the equations.}

\begin{document} 
\maketitle
\flushbottom

\section{Introduction}

During the last decades our knowledge of gravity has extremely been improved with general relativity (GR) remaining our best classical theory to describe it. Some inconsistencies or debates in the cosmology community rise from time to time, such as the Hubble constant tension nowadays, but they are in no way a direct test of GR. All direct tests are consistent with the theory, see e.g. \cite{Will:2014kxa}.
On the other hand, GR is a classical theory and therefore partial. It is for example geodesically incomplete for most of its solutions \cite{Hawking:1973uf} like black holes which contain spacetime singularity. Also, some theoretical arguments challenge our knowledge in cosmology and the existence of the cosmological constant. Recently, a lot of attention has focused on the String Swampland \cite{Vafa:2005ui,Palti:2019pca} which rejects any de Sitter solution \cite{Agrawal:2018own} and therefore the existence of the cosmological constant. The origin of the recent acceleration of the universe could be due to a scalar field. This would be a very interesting promotion of the relevance of scalar fields in the dynamics of the universe. At the same time, many studies try to see if scalar fields could be locally observed, and what are their effects around black holes. For example, we can ask if we could observe any deviation from Kerr black hole by measurement of quasi-normal modes \cite{Giesler:2019uxc}. Even if the answer is, at present time, negative, quintessence and their extensions with non-standard kinetic terms, attracted considerable interest. It is rather common for effective field theories to have scalar fields and non-linear terms, such as e.g. the ones originated from D-branes models \cite{Garousi:2000tr,Bergshoeff:2000dq,Sen:2002in}.

The first non-linear model of this type has been proposed in 1934 by Born and Infeld \cite{Born:1934gh} with a non-linear electromagnetic field to avoid the infinite self-energy of the electron in classical electrodynamics. For our concern in this paper, the models originated in cosmology in the context of inflation \cite{ArmendarizPicon:1999rj} and later adapted to dark energy \cite{ArmendarizPicon:2000ah}. These models can also be used to describe dark matter \cite{Sen:2002in,ArmendarizPicon:2005nz}. This last approach permits us to approximate dark matter by a non-linear quintessence field and therefore identify the possible effects of dark matter on gravitational collapse. 

The numerical study of spherical gravitational collapse has a long history which began with the work of the collapse of ideal fluid spheres with an equation of state $P=2\rho/3$ \cite{May:1966zz}. They found that collapse could lead to the formation of a black hole or a bounce according to initial conditions. Many new codes have been later developed with a focus on applications to realistic stellar collapse. But there has also been considerable interest into more theoretical problems such as critical phenomena. Choptuik \cite{Choptuik:1992jv} has shown that if $p$ is a parameter describing some aspect of the initial distribution of scalar field energy, there exists a critical value $p^\star$ which denotes the threshold of black hole formation. For $p<p^\star$, the scalar field disperses to infinity while for $p>p^\star$ black hole forms. In the supercritical regime, meaning for $p>p^\star$ but very close to the threshold, a universal behavior appears (i.e. independent of the initial data) relating the mass $M$ of black holes to a universal scaling behavior
\begin{align}
M\propto (p-p^\star)^{\gamma}~,~~~\gamma\simeq 0.37
\end{align}
This solution has been repeatedly verified, also by using a fully 3D code \cite{Healy:2013xia}. But as in critical phenomena, there exist classes of universality. Adding a mass term to the theory \cite{Brady:1997fj}, which introduces a length scale, produces also a universal behavior but with a different scaling parameter $\gamma$. See also studies with a massive complex scalar field \cite{Hawley:2000dt}, with radiation fluid \cite{Evans:1994pj} or with extra dimensions \cite{Sorkin:2005vz}.

In this paper, we study a natural extension of the work performed by Choptuik, by studying models known as K-essence. We will consider the generic theory and summarize the various conditions for the viability of these models at classical as well as quantum level. We will derive the characteristics for these models and therefore their hyperbolicity. In section 3, using a spherically symmetric spacetime we will obtain the constraints and evolution equations. For numerical purposes, we will assume a particular K-essence model which will be studied in the weak as well as the strong gravitational regime before conclusions.
    
\section{K-essence}

Let us consider the following action 
\begin{align}
\mathcal{S}=\int {\rm d}^4x \sqrt{-g}\Bigl[\frac{R}{2}+K(\phi,X)\Bigr]
\end{align}
where $\phi$ is a scalar field representing the matter sector, $X=-\frac{1}{2}\partial_\mu\phi\partial^\mu\phi$ is the canonical kinetic term and $K$ is a generic function of the scalar field and the kinetic term\footnote{Note that our metric signature is $(-,+,+,+)$}. Considering only the sub-class of scalar-tensor models of gravity, more generalized extensions have been constructed, from Galileons \cite{Nicolis:2008in}, to Horndeski \cite{Deffayet:2009wt,Kobayashi:2011nu,Horndeski:1974wa} to beyond Horndeski \cite{Zumalacarregui:2013pma,Gleyzes:2014dya,Gleyzes:2014qga}. Even if these models seem to have been finely constructed, they appear to have a well-posed Cauchy problem only in high symmetrical backgrounds such as Friedmann or spherically symmetric spacetimes. But generically, they suffer from a major problem \cite{Papallo:2017qvl}. The equations of motion are not strongly hyperbolic for most Horndeski models except K-essence which arise as the most legitimate sub-class of scalar-tensor theories.

In this paper, we will consider models where the action is a function of $X$ only, it inherits an additional shift symmetry i.e. an  invariance under constant translation in field space, $\phi \rightarrow \phi +c$, for any constant $c$. It is important to notice that demanding the existence of  stationary configurations requires shift symmetry \cite{Akhoury:2008nn}, sometimes after a field redefinition. This restriction makes the model equivalent to a perfect fluid with no vorticity. In fact, the variation of the action in shift symmetry models gives\footnote{We consider $G=c=1$} $G_{\mu\nu}=8 \pi T_{\mu\nu}$, where the energy-momentum tensor is defined as
\begin{align}
T_{\mu\nu}=K_{,X}\partial_\mu\phi \partial_\nu\phi+g_{\mu\nu}K
\end{align}
It is well known that this stress-energy tensor can be put in a hydrodynamical language,
\begin{align}
T_{\mu\nu}=(\rho+P)u_\mu u_\nu +P g_{\mu\nu}
\label{eq:Tmunu}
\end{align}
where we define an effective four-velocity $u_\mu=\partial_\mu\phi/\sqrt{2X}$, density $\rho=2XK_{,X}-K$ and pressure $P=K$. We see also that the pressure is a function of the energy density only, $P=P(\rho)$. Therefore, choosing an action is equivalent to specifying an equation of state (EoS) for the equivalent hydrodynamical model. For example, considering $K=(\alpha X^{1/2\beta}-A)^\beta$, the EoS is $P=A\rho^{(\beta-1)/\beta}$, a polytropic law similar to various models describing neutron stars (without the anisotropic stress tensor).
Notice also that $K(X)$ models can be related to canonical complex scalar field theories \cite{Babichev:2018twg} where the potential of the complex scalar field is defined by $K(X)$. More than an equivalence between these theories, it is an extension or UV completion. In fact, the complex scalar field can be seen as a theory with 2 real scalar fields with an $O(2)$ symmetry. It's only when one of these fields is frozen, that the model reduces to K-essence.

For shift-symmetric K-essence models, the speed of sound for small perturbations around a given background coincide with the usual definition of the sound speed for the perfect fluid \cite{Garriga:1999vw},
\begin{align}
c_s^2=\frac{K_{,X}}{K_{,X}+2XK_{,XX}}\equiv \frac{\partial P}{\partial \rho}
\end{align}
The variation of the action wrt the scalar field gives
\begin{align}
\label{eq:KG}
\nabla_\mu(K_{,X}\nabla^\mu\phi)=\tilde{g}^{\mu \nu}\nabla_{\mu\nu}\phi = 0
\end{align}
where the effective metric is defined as
\begin{align}
\label{eq:gtilde}
\tilde{g}^{\mu\nu}=g^{\mu\nu} K_{,X}-K_{,XX}\partial^\mu\phi\partial^\nu\phi
\end{align}
or the inverse metric
\begin{align}
\tilde{g}_{\mu\nu}=\frac{1}{K_{,X}} g_{\mu\nu}+c_s^2\frac{K_{,XX}}{K_{,X}^2} \partial_\mu\phi\partial_\nu\phi
\end{align}
A theorem due to Leray \cite{Wald:1984rg} proves that the generalized Klein-Gordon equation has a well posed Cauchy problem if the metric is Lorentzian which has been proved to be equivalent to $c_s^2>0$ \cite{ArmendarizPicon:2005nz}. This condition is often referred to as the classical condition. On the other hand, a stronger condition related to the Hamiltonian of field perturbations to be positive definite (in cosmological context) implies $K_{,X}>0$ and $K_{,X}+2XK_{,XX}>0$ \cite{ArkaniHamed:2003uy}, this condition is often dubbed in the literature as the quantum stability condition, because if other sectors such as gravity or standard model particles couple to scalar field described by an unbounded Hamiltonian from below, it would create states of atoms which never decay if excited, a situation never observed in nature (a more careful discussion is proposed in \cite{Babichev:2018uiw}).

K-essence models are considered as an effective low energy description of some more fundamental theory. Therefore, we should impose it to be consistent with basic requirements of quantum field theory, such as Lorentz
invariance, unitarity, analyticity... \cite{Adams:2006sv,Nicolis:2009qm}. Considering the tree-level scattering amplitude, between two massive particles on a flat background, these restrictions impose $K_{,XX}>0$ \cite{Melville:2019wyy}. All these conditions imply non-superluminal propagation $(c_s^2<1)$. 

In summary, we consider models of gravity defined by a shift-symmetric K-essence action with conditions $K_{,X}>0$ and $K_{,XX}>0$. Notice that gravitational collapse for such models have been previously considered \cite{Leonard:2011ce} but with models violating one of these conditions. For there models, a sonic horizon ($\tilde{g}^{rr}=0$) could be defined inside the luminal apparent horizon ($g^{rr}=0$) during gravitational collapse because the model allow superluminal propagation. As it has been shown for the first time in \cite{Babichev:2006vx}, perturbations of the scalar field could escape from the inside of the black hole defined by the luminal apparent horizon without violating causality, because they emerge outside of the sonic horizon. By imposing our conditions, we exclude these situations. The sonic horizon will be always larger than the luminal apparent horizon and might merge together in the future.

Notice that these conditions imply that null energy conditions will not be violated, which turned out to be central in singularity theorems. The weak energy condition plays an important role, it implies that matter has always a non diverging effect on congruences of null geodesics. It has been very influential, e.g. the area theorem proved by Hawking \cite{Hawking:1973qla,Hawking:1973uf} states that if matter satisfies the null convergence condition
or equivalently in general relativity the null energy condition\footnotetext{For every null vector $n^\mu$, the null convergence condition is defined as $R_{\mu\nu}n^\mu n^\nu\geq 0$ while $T_{\mu\nu}n^\mu n^\nu\geq 0$ defines the null energy condition}, the area of the black hole event horizon can never decrease, statement very similar to the second law of thermodynamics \cite{Ashtekar:2004cn}.

\section{Model and equations}

In polar-areal coordinates \cite{Choptuik:1992jv}, the metric takes the form
\begin{align}
{\rm d}s^2 = -\alpha(t,r)^2 {\rm d}t^2+a(t,r)^2 {\rm d}r^2+r^2({\rm d}\theta^2+\sin^2\theta {\rm d}\phi^2).
\end{align}
If simpler, this choice is not the most appropriate because it is valid until it forms a trapped surface. Unfortunately, as we will see a trapped surface associated with the effective metric might form before the normal trapped surface, which will break the numerical evolution before the formation of the black hole. The sonic horizon forms when $\tilde{g}^{rr}=0$ which always forms before the apparent horizon defined by $g^{rr}=0$, which can be seen easily from eq.(\ref{eq:gtilde}) if we assume the conditions $K_{,X}>0$ and $K_{,XX}>0$. Also, in general, to solve the gravitational collapse we foliate the spacetime with spacelike hypersurfaces, a  breakdown of the evolution occurs when the hypersurface of constant time becomes null, i.e.
\begin{align}
g^{\mu\nu}\nabla_\mu t \nabla_\nu t = 0 ~~~|| ~~~\tilde{g}^{\mu\nu}\nabla_\mu t \nabla_\nu t = 0
\end{align}
which implies $g^{00}=0$ or $\tilde{g}^{00}=0$. Using eq.(\ref{eq:gtilde}), it is also trivial to see that following our conditions, if $g^{00}=0$ occurs, it will always happen before $\tilde{g}^{00}$ vanishes. Therefore, we conclude that our numerical evolution will fail if $g^{00}=0$ or $\tilde{g}^{rr}=0$. Even if this coordinate system is not the most appropriate, we will be able to deduce some interesting results. 

In order to reduce the K-essence field equation of motion to a system of first-order PDEs, we define two auxiliary fields \cite{Choptuik:1992jv}
\begin{align}
\Phi(t,r) &\equiv \partial_r \phi(t,r)\\
\Pi(t,r) &\equiv \frac{a(t,r)}{\alpha(t,r)}\partial_t\phi(t,r)
\end{align}
The $tt$- and $rr$-components of Einstein equation gives
\begin{align}
\label{eq:constraint1}
\frac{a'}{a}+\frac{a^2-1}{2 r} &=4\pi r \Bigl(K_{,X}\Pi^2-a^2 K\Bigr)\\
\label{eq:constraint2}
\frac{\alpha'}{\alpha}-\frac{a^2-1}{2 r} &=4\pi r \Bigl(K_{,X}\Phi^2+a^2 K\Bigr)\\
\text{with}~~X &=\frac{1}{2a^2}\Bigl(\Pi^2-\Phi^2\Bigr)
\label{eq:X}
\end{align}
These equations contain no time derivatives, so they are constraints, they must be satisfied at each time. These are the same equations than the Hamiltonian and Momentum constraints obtained after a $3+1$ decomposition of Einstein equations. The dynamics of the system is given by the definition of the auxiliary fields which imply
\begin{align}
\label{eq:Phi}
\dot\Phi = \partial_r\Bigl(\frac{\alpha}{a}\Pi\Bigr)
\end{align}
The second evolution equation is given by the generalized Klein-Gordon equation (\ref{eq:KG}) and the $tr$-component of the Einstein equation
\begin{align}
\label{eq:Pi}
\Bigl(K_{,X}+K_{,XX}\frac{\Pi^2}{a^2}\Bigr)\dot\Pi = \frac{1}{r^2}\partial_r\Bigl(r^2\frac{\alpha}{a}\Phi K_{,X}\Bigr)+8\pi r \frac{\alpha}{a}\Phi \Pi^2 X K_{,X} K_{,XX}+K_{,XX}\frac{\Phi \Pi}{a^2}\partial_r\Bigl(\frac{\alpha}{a}\Pi\Bigr)
\end{align}
For numerical purposes, we need to define a particular model. We choose the simplest extension of quintessence, which fulfill the previous conditions
\begin{align}
K(X)=X+\beta X^2
\end{align}
The constant $\beta$ could take any value but as mentioned previously, we need to impose the condition $K_{,XX}>0$ which suggests $\beta>0$. This condition which implies a standard UV completion of the theory, turns out to be also related to hyperbolicity of the equations and therefore causality. Maybe a deep relation could be obtained between hyperbolicity of classical equations and standard Wilsonian field theory description of the quantum version.

To integrate our system of equations, we define an initial profile of the field $\phi(t=0,r)$ implying $\Phi(0,r)$ to which we add a second initial condition $\Pi(0,r)=0$. These two functions are sufficient to integrate the constraint equations (\ref{eq:constraint1},\ref{eq:constraint2}) by assuming boundary conditions. Regularity of the system imposes $a(t,r=0)=1$, and without loss of generality we choose $\alpha(t,r=0)=1$ which corresponds to choosing the time coordinate at $r=0$ to be the proper time. A change in this value corresponds to a trivial rescaling of the time coordinate and would have no physical consequences. Fourth order Runge-Kutta (RK4) method is used to integrate these constraint equations. These values of $(\alpha,a)$ are then entered into the evolution equations (\ref{eq:Phi},\ref{eq:Pi}) to find $(\Phi,\Pi)$ at the next time step with RK4. This process is repeated until the scalar field disperses to infinity and forms flat spacetime or until forms an apparent horizon. The two families of initial data that we adopt are
\begin{align}
&\text{Family A:}~~~\phi(0,r)=\phi_0 r^3 e^{-(\frac{r-r_0}{d})^q}\\
&\text{Family B:}~~~\phi(0,r)=\phi_0 \tanh\frac{r-r_0}{d}
\end{align}
where ($\phi_0,r_0,d,q$) are constants. For each family of initial conditions, we keep only $\phi_0$ as a free parameter, the others are fixed to $(r_0=20,q=2,d=3)$. The system is evolved between $r=10^{-50}$ and $r=50$ from $t=0$ until it forms an apparent horizon ($r_H$) featuring fixed mesh refinement. The r-spacing $\Delta r$ varies from the finest value near the origin to larger values of $r$ in 5 different sectors. Near the origin and until $r_H$ (which is approximately determined in a first run) the resolution is $\Delta r=10^{-4}$, this r-spacing is progressively increased 4 times until it reaches $10^{-2}$ at larger $r$. The time resolution is also fixed but satisfies the Courant-Friedrichs-Lewy condition $\Delta t=\Delta r/5$ where $\Delta r$ takes 5 different values as defined by each sector (from $\Delta r=10^{-4}$ to $\Delta r=10^{-2}$). All results are verified by modifying the resolution in space and time and for most of them we checked with a fixed mesh of $\Delta r=10^{-4}$ in all space. All results presented in the paper are stable under these tests.

We should emphasize that the energy-momentum tensor defined in eq.(\ref{eq:Tmunu}) describe an observer with four-velocity $u_\mu=\partial_\mu\phi/\sqrt{2X}$ and therefore in general case with a radial velocity. For that observer, the energy density could be negative. In fact, even for the simplest case where $K=X$, we would have $\rho=X$, which from eq.(\ref{eq:X}) and the initial condition $\Pi=0$ gives $\rho=-\Phi^2/2a^2$ which is at initial time, negative. In order to define the energy density measured by a static observer at position $r$, we would need a new four-velocity $n^\mu=(1/\alpha,0,0,0)$ and therefore for this observer an energy density 
\begin{align}
\bar\rho=T_{\mu\nu}n^\mu n^\nu=2XK_{,X}+\frac{\Phi^2}{\Pi^2-\Phi^2}(K+2XK_{,X})
\end{align}
which at $t=0$ gives $\bar\rho=-K=-X>0$ for $K=X$.

\section{Characteristics}
Following standard textbooks \cite{Courant:1962}, we compute the characteristic structure of our system (see also \cite{Ripley:2019hxt}). It is sufficient to analyze the evolution equations of $(\Phi,\Pi)$ defined in eqs.(\ref{eq:Phi},\ref{eq:Pi}) in which we need to replace $(\alpha',a')$ from the constraint equations to reach a system of 2 equations of the following form $E^{(i)}[\alpha,a,w^{(j)},\partial_r w^{(j)}, \partial_t w^{(j)}]=0$ where $w^{(1)}=\Phi$ and $w^{(2)}=\Pi$. We introduce the principal symbol 
\begin{align}
P^i_j(\xi_a)\equiv \frac{\delta E^{(i)}}{\delta (\partial_a w^{(j)})}\xi_a
\end{align}
where $(\xi_t,\xi_r)$ define the characteristic covector. By solving the characteristic equation defined by det$[P^i_j(\xi_a)]=0$, we deduce the characteristic speed as $c=-\xi_t/\xi_r$.
\begin{align}
\label{char}
c_\pm = \frac{-K_{,XX}a\alpha \Pi \Phi\pm a^3\alpha \sqrt{K_{,X}(K_{,X}+2XK_{,XX})}}{K_{,X}a^4+K_{,XX}a^2\Pi^2}
\end{align}
In the case of a canonical scalar field, $K=X$, we obtain $c_\pm = \pm \alpha/a$ which reduce to the characteristic speeds of GR and the equation is always hyperbolic. In the generic case, the sign of $K_{,X}(K_{,X}+2XK_{,XX})$ defines the character of the system. If positive, it is hyperbolic, when negative it is elliptic and if $K_{,X}(K_{,X}+2XK_{,XX})=0$ it is parabolic. 
Notice that, this condition is similar to $\det \tilde g =0$ or equivalently to the sign of the eigenvalues of the effective metric $\tilde g$. Even if we impose the condition  $\beta>0$, the hyperbolicity of the equations can be lost after some time of evolution. In fact, the condition $K_{,X}(K_{,X}+2XK_{,XX})\equiv (1+2\beta X)(1+6\beta X)>0$ can be violated if $-1<2\beta X<-1/3$ (for any sign of $\beta$). Because we imposed $\beta>0$, this condition is violated if $X$ is sufficiently negative which translates into a model developing large enough space-like gradients ($\Phi>\Pi$ from eq.(\ref{eq:X}))\footnote{We thank the anonymous referee for suggesting this possibility}. This behavior was already predicted in \cite{ArmendarizPicon:2005nz, Rendall:2005fv}. But remained to know if for generic initial conditions, large space-like gradients of the solution could form. This was recently proved numerically to happen in \cite{Bernard:2019fjb}. But $\beta >0$ remains a better option which allows some initial conditions to develop without any loss of hyperbolicity, contrary to $\beta<0$ case.

\section{Numerical results}
\subsection{Weak field regime}
\begin{figure}
\includegraphics[scale=0.3]{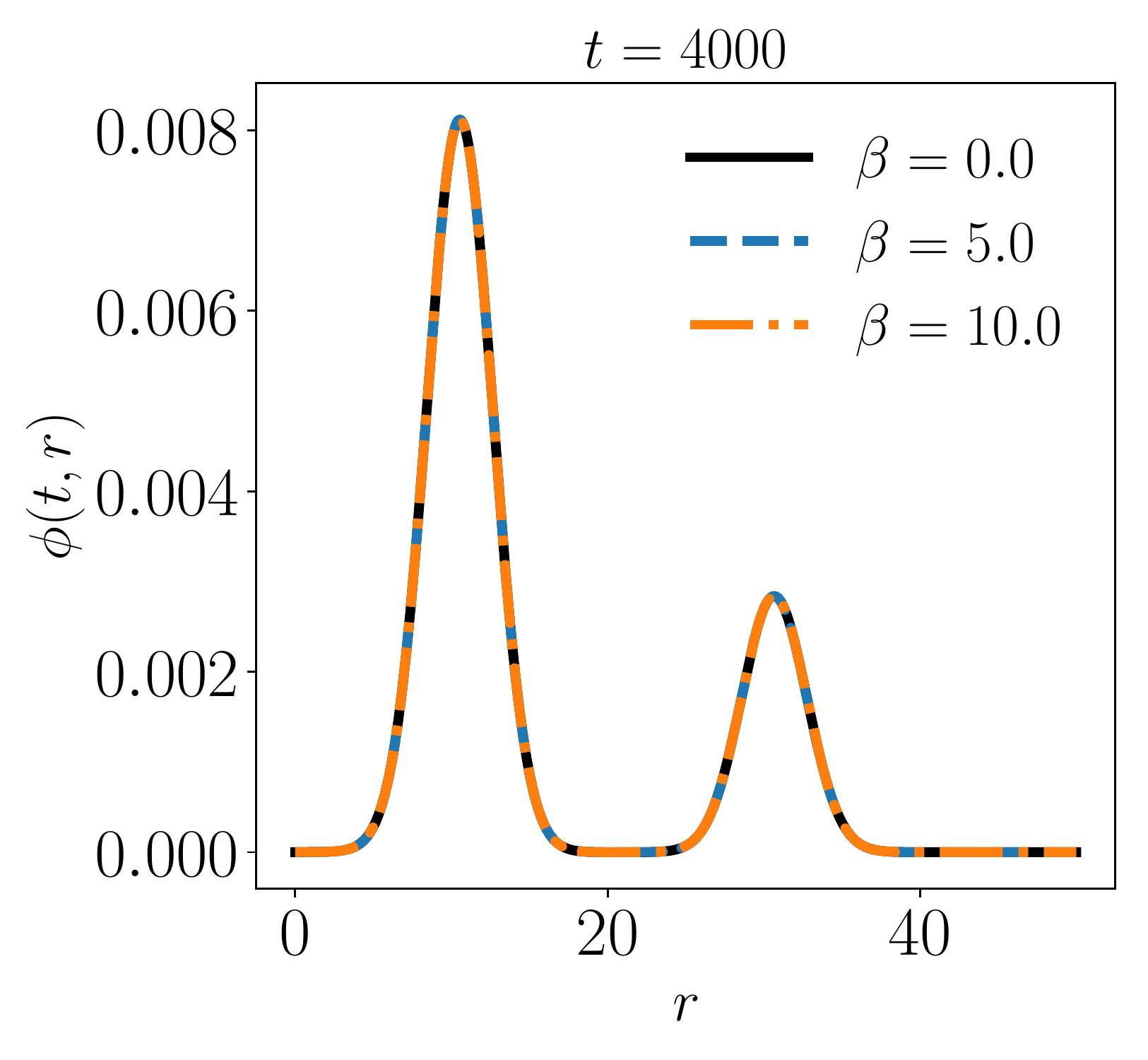}	
\includegraphics[scale=0.3]{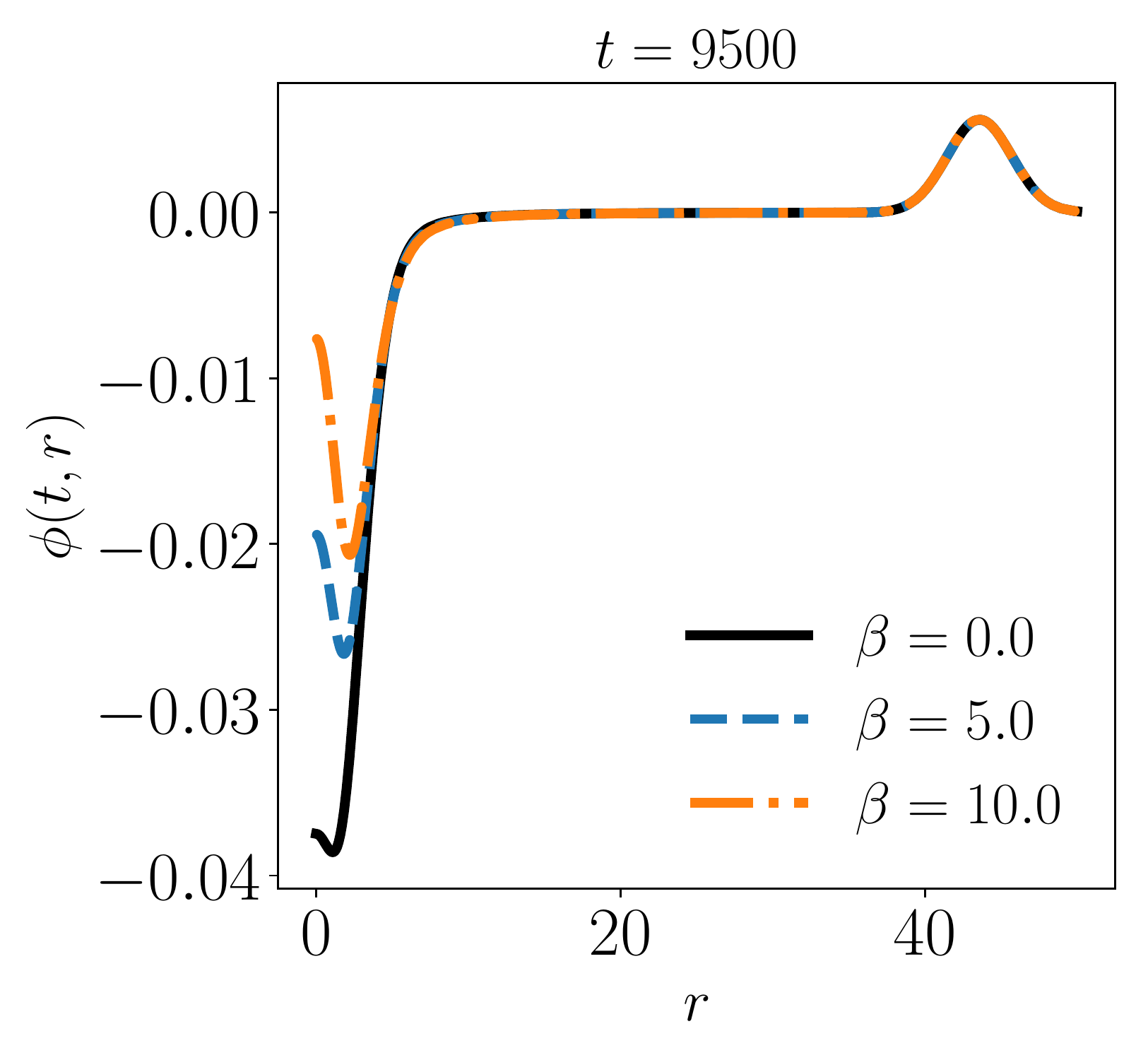}	
\includegraphics[scale=0.3]{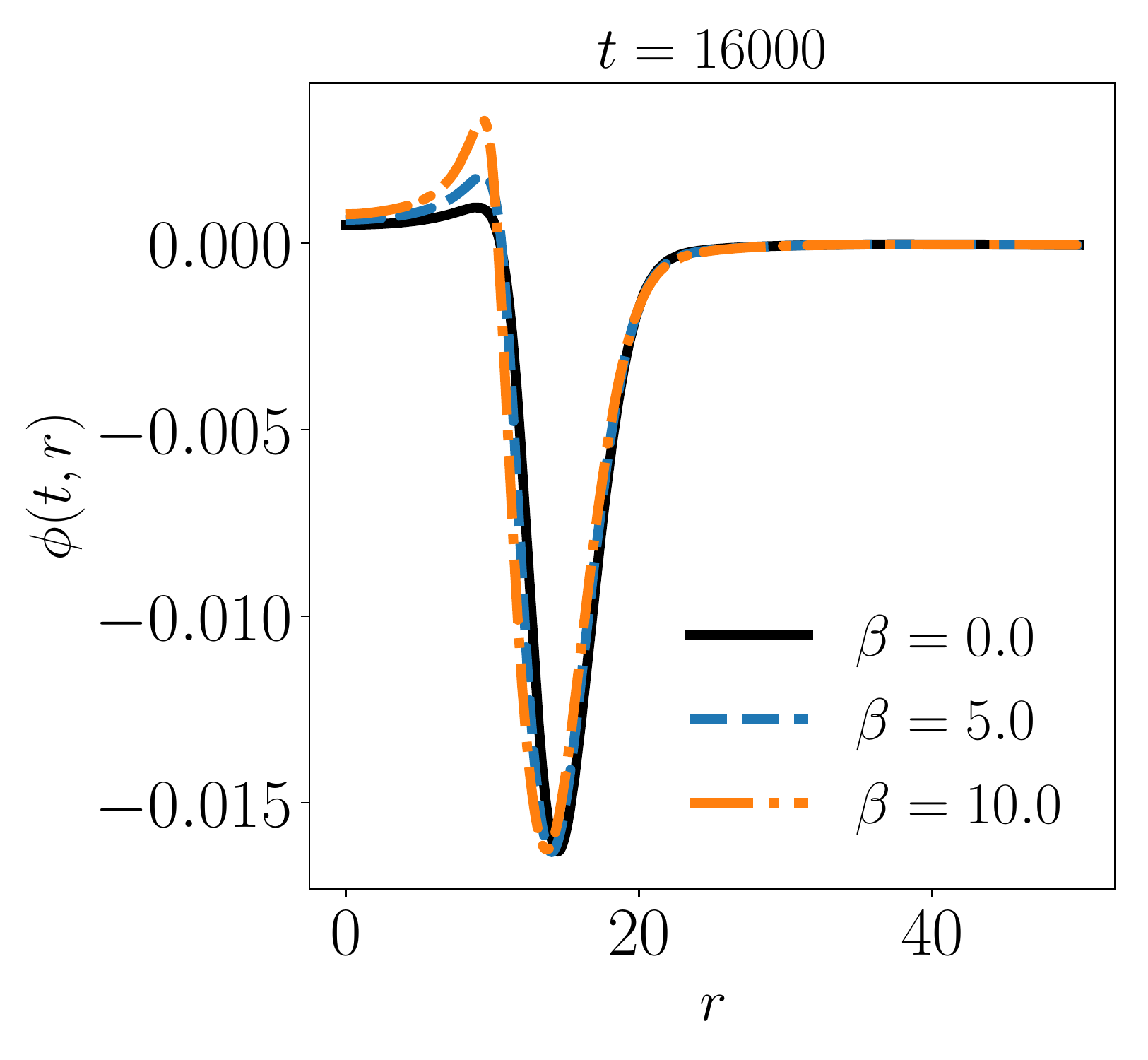}	
\includegraphics[scale=0.3]{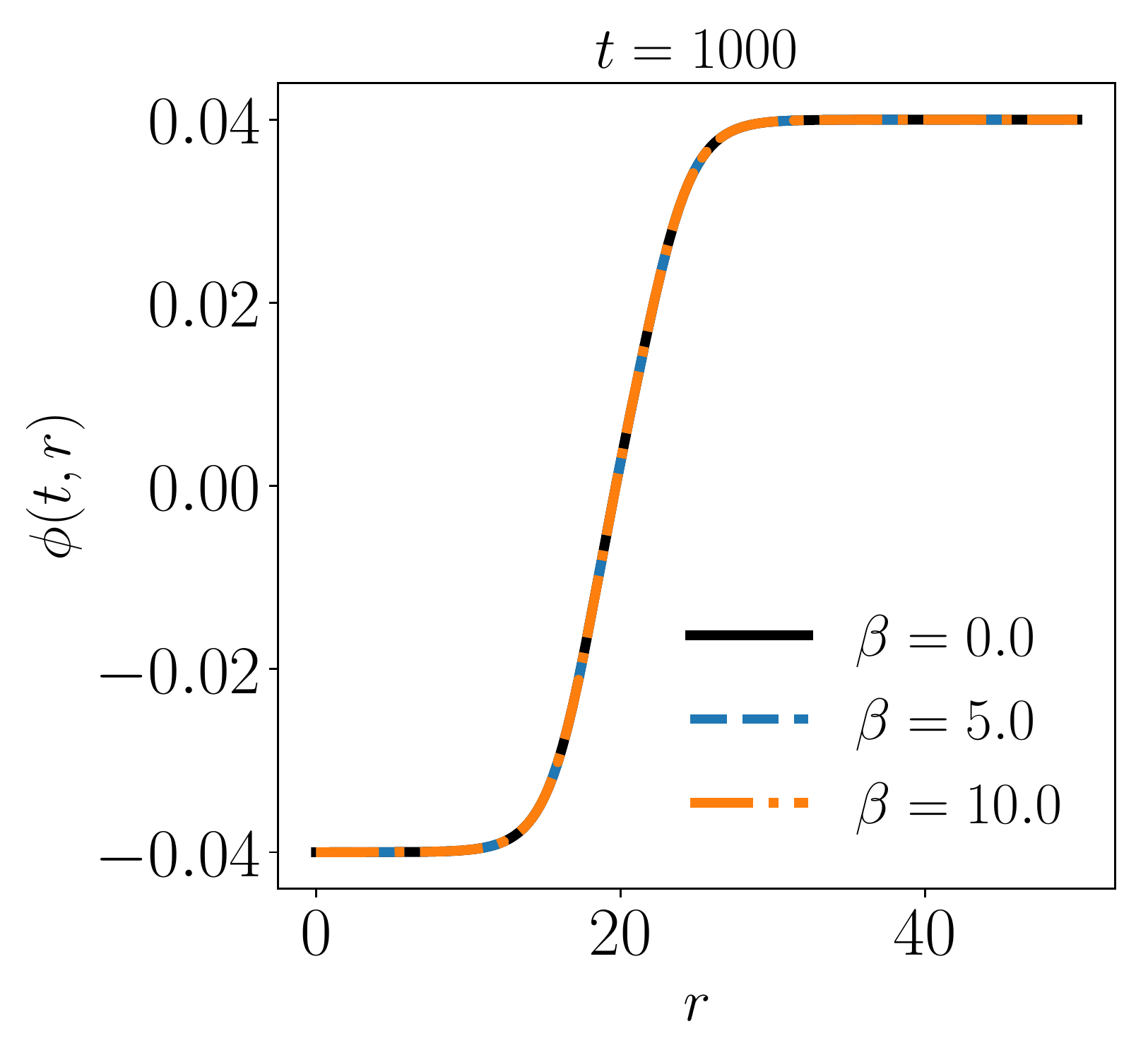}	
\includegraphics[scale=0.3]{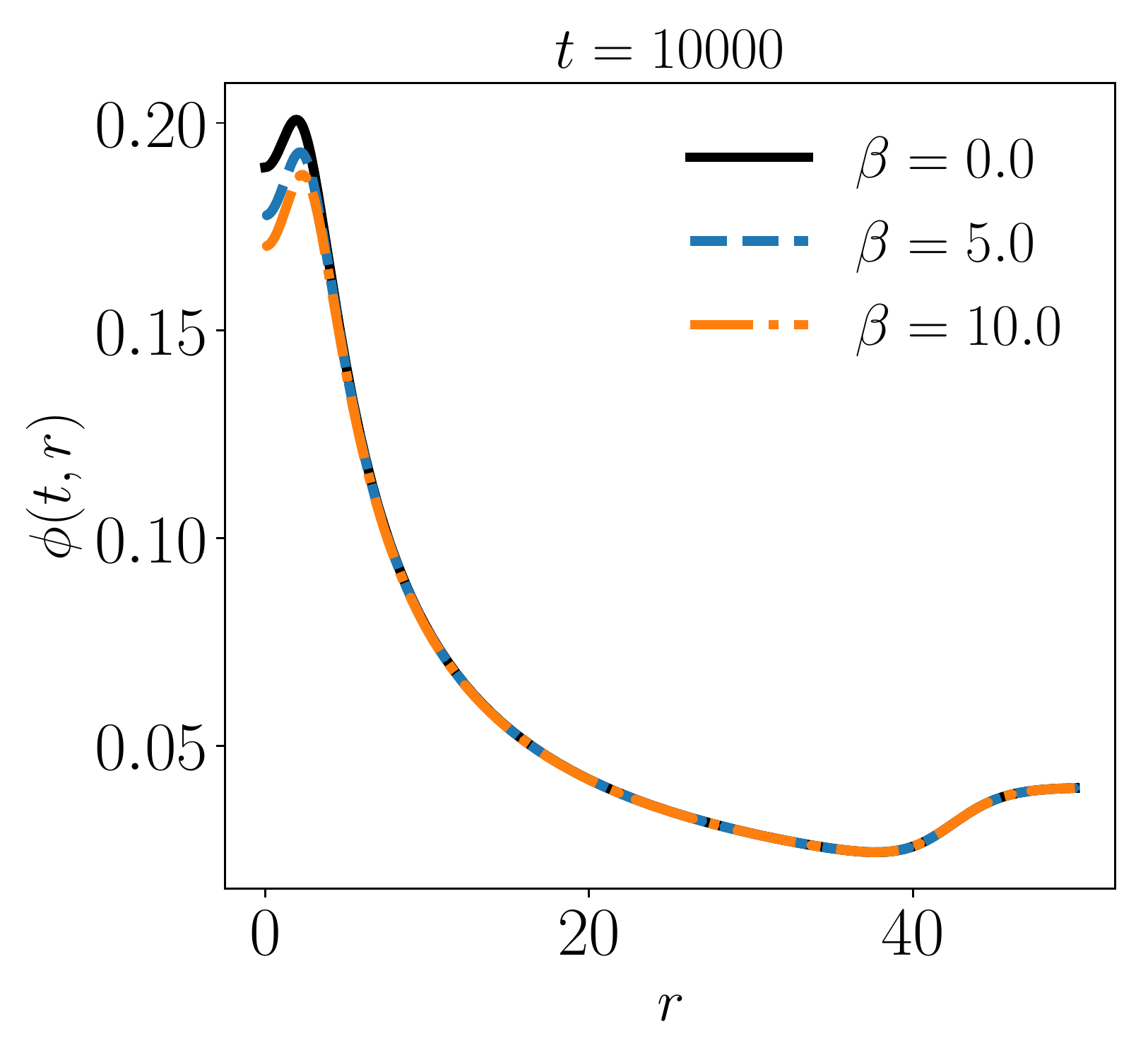}	
\includegraphics[scale=0.3]{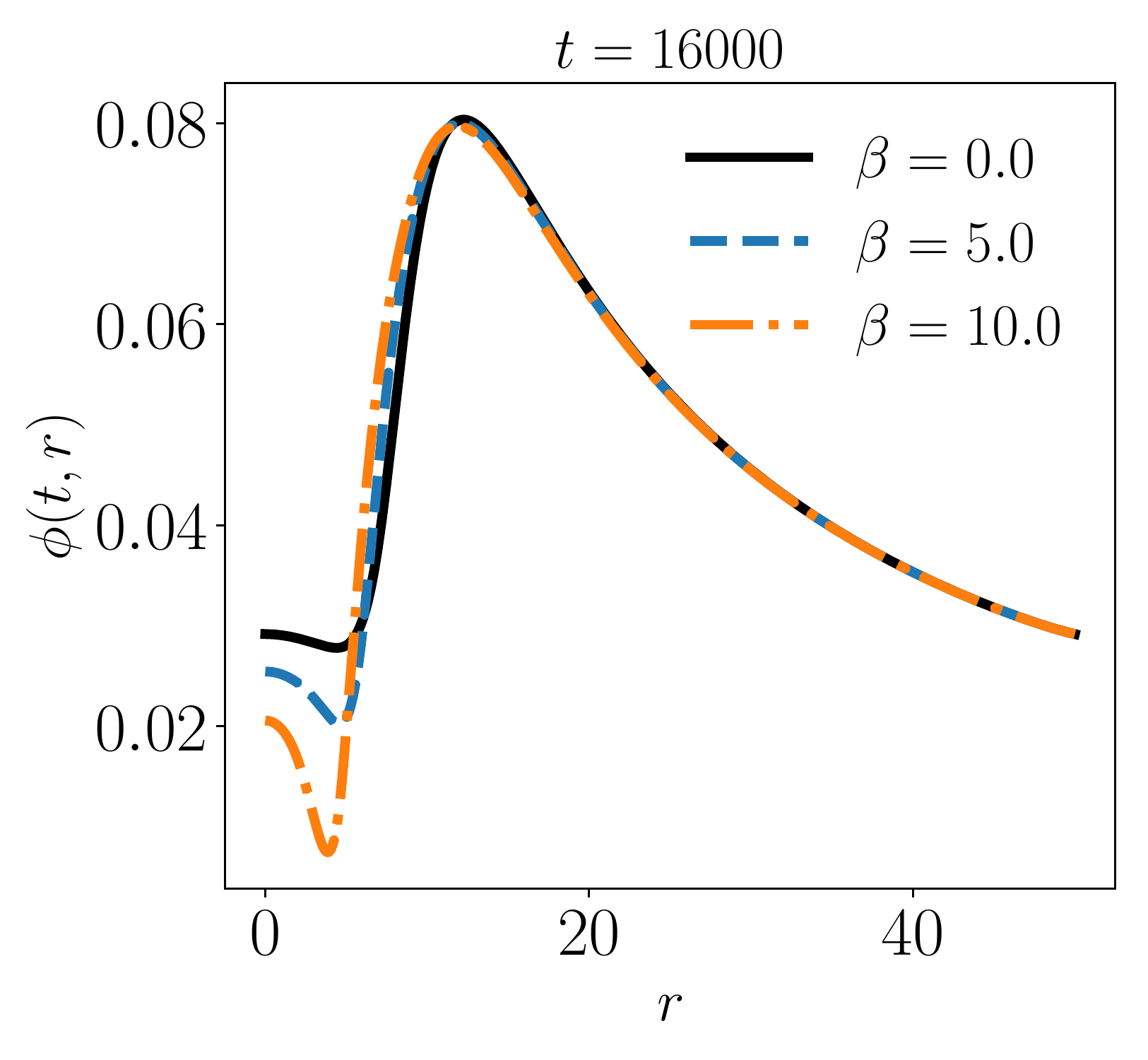}	
\caption{Scalar field profile $\phi$ in the weak field  regime, for $\beta = 0.0$ (solid line),  $\beta = 5.0$ (dashed line) and $\beta = 10.0$ (dash-dotted line). Upper panel is for Gaussian family type of initial conditions (Family A) while the bottom panel represents family B. \label{Fig:weak}}
\end{figure}

In the weak field regime, when $\phi_0\ll 1$, the scalar field bounces at the origin $(r=0)$ and then disperses to infinity. We see in Fig.(\ref{Fig:weak}) this behavior for 3 different time\footnote{Time is the iteration step and not proper time at $r=0$} and for 3 different values of $\beta=\{0,5,10\}$ for the 2 families of initial conditions. For $t=4000$ for family A and $t=1000$ for family B, the field is yet collapsing. Around $t=9500$ for Gaussian initial conditions and $t=10000$ for the second family of initial conditions the field bounces at $r=0$ and finally at a later time, the field disperses to infinity. We see that the parameter gives a small variation to the dynamics of the scalar field because of the weak regime studied in this section. Even if we can notice that for larger values of $\beta$, the field takes more time to bounce and therefore reaches infinity a bit later compared to $\beta=0$. Notice that taking larger values of the constant $\beta$ increases the mass of the spacetime and therefore we get closer to the threshold of black hole formation.

\subsection{Strong field regime}
\begin{figure}
\includegraphics[scale=0.34]{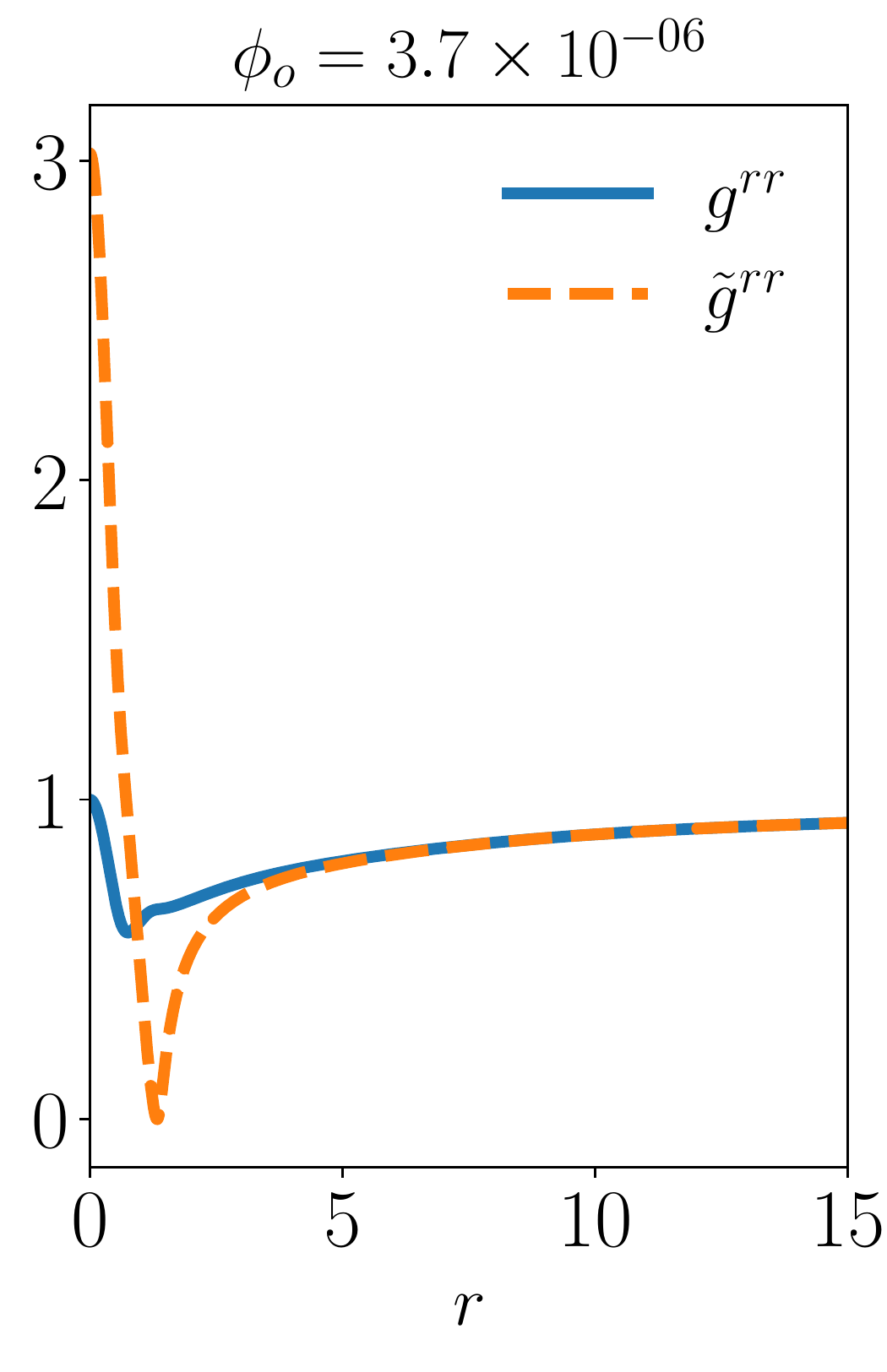}	
\includegraphics[scale=0.34]{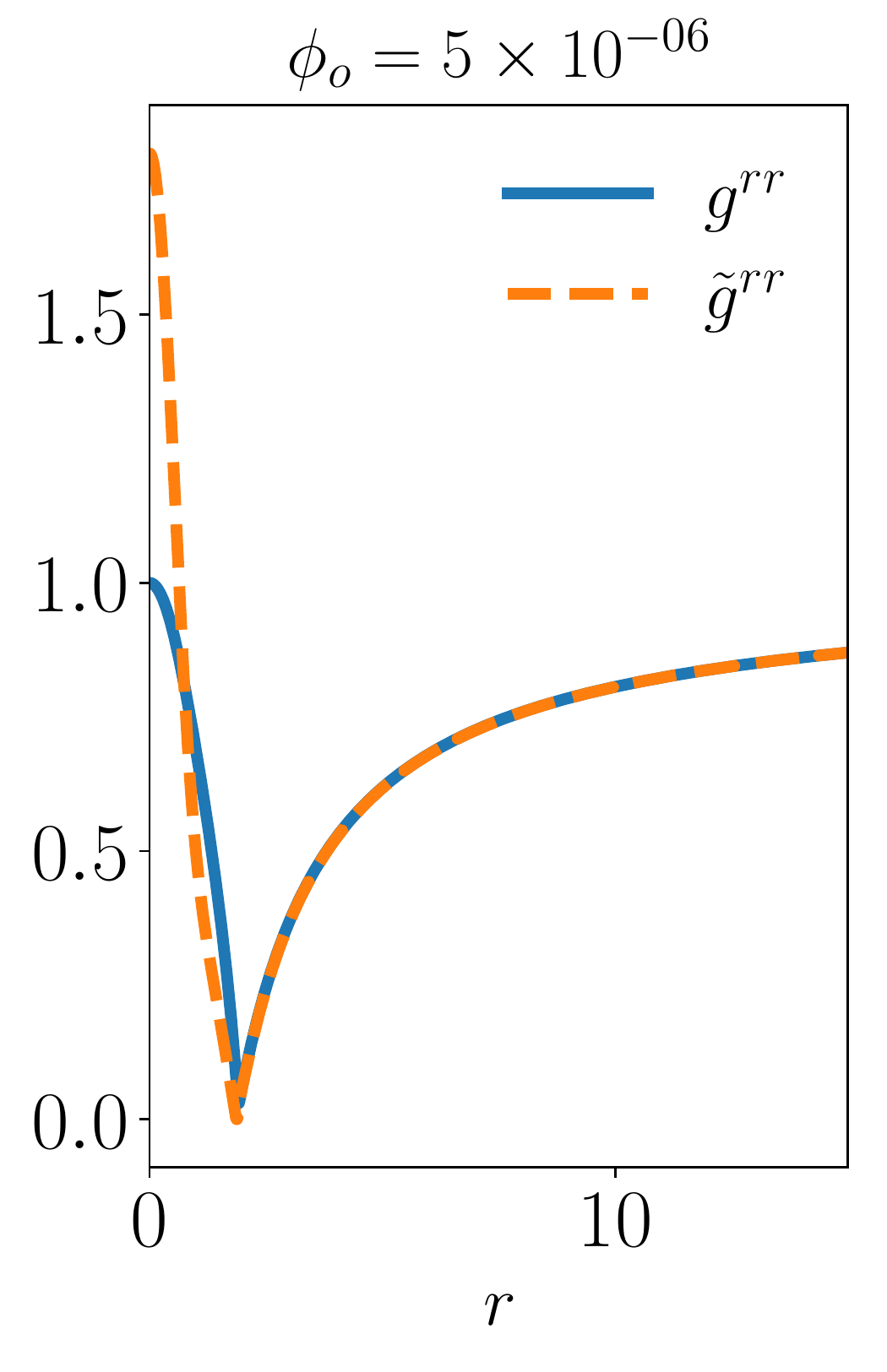}		
\includegraphics[scale=0.34]{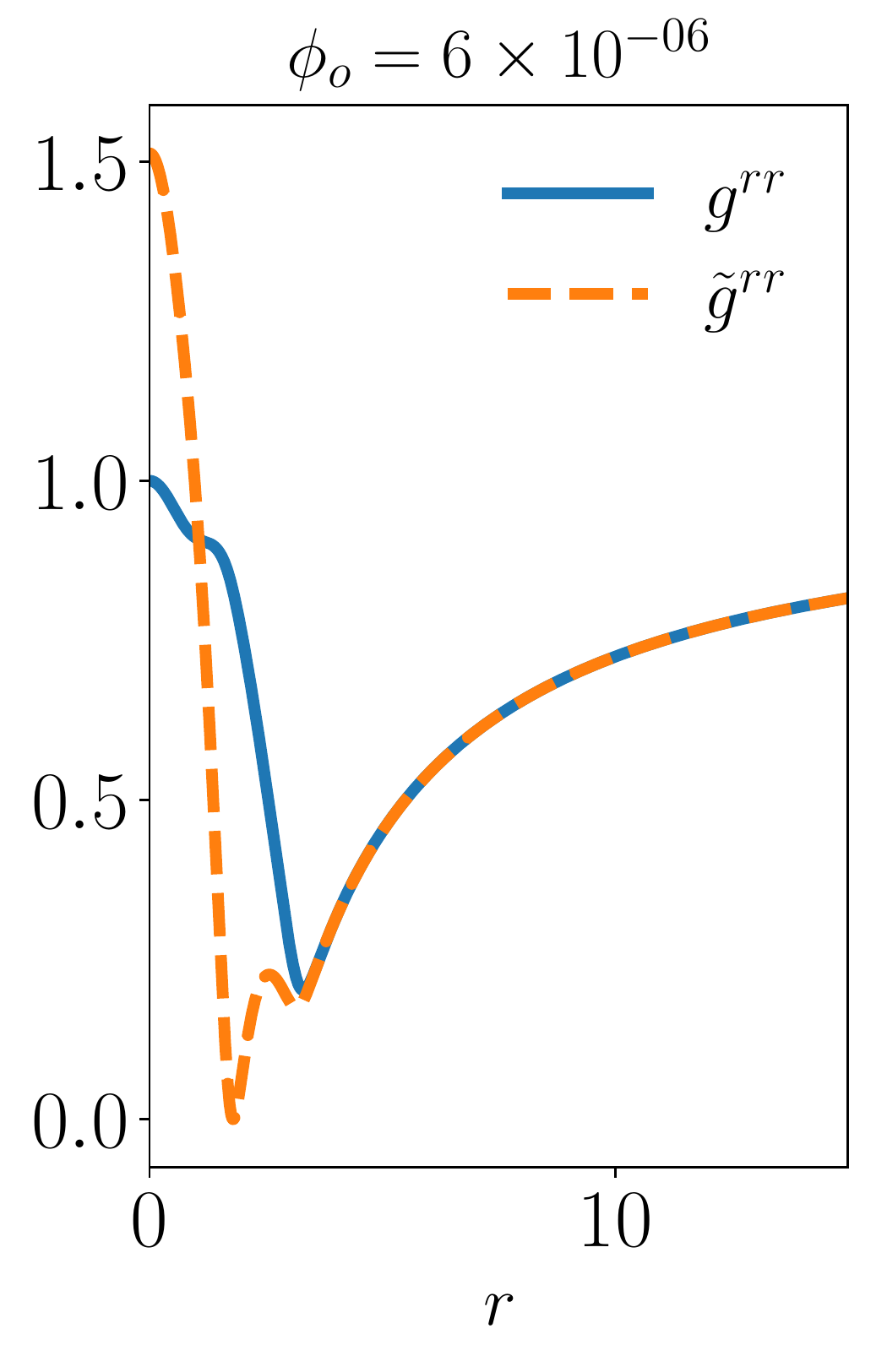}		
\includegraphics[scale=0.34]{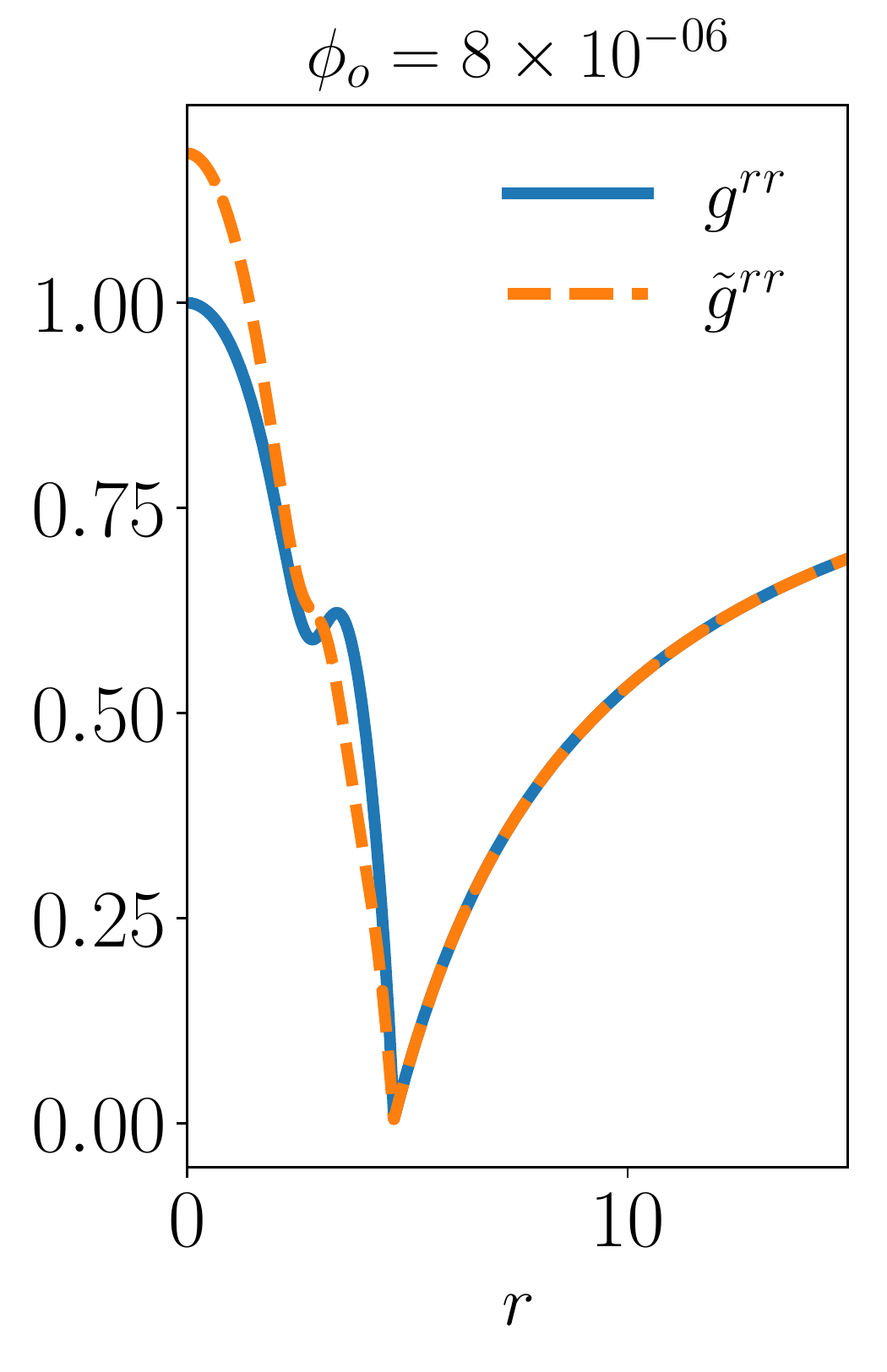}		
\caption{Last time evolution before formation of a horizon. The event horizon forms when $g^{rr}=0$ and the sonic horizon when $\tilde{g}^{rr}=0$. For $\phi_0=3.7~ 10^{-6}$ and $\phi_0=6 ~10^{-6}$ forms a sonic horizon while the event horizon is not yet formed, while for $\phi_0=5~10^{-6}$ and $\phi_0=8~10^{-6}$ both horizons form at the same time within our numerical precision.
\label{Fig:4regimes}}
\end{figure}

We know that when the initial mass of the scalar field added in the spacetime is large, the collapse of that field produces a black hole. In this paper, we run simulations varying the amplitude of the K-essence scalar field $\phi_0$ from a value in the weak limit regime to a value where the final result of the evolution is the creation of an event horizon (collapse of the metric) or a sonic horizon (collapse of the effective metric). As we described previously, since our metric is not horizon penetrating, we cannot evolve the spacetime beyond the formation of any horizon. We found 4 different regimes. For $\beta=5$, we found situations described in Fig.(\ref{Fig:4regimes}) but which are generic for any value of $\beta \neq 0$. We found
 
\begin{itemize}
\item For an amplitude of the scalar field small, $\phi_0=3.7~ 10^{-6}$ for $\beta=5$, we observe the formation of a sonic horizon without being able to continue the simulation to know if a black hole forms. But it is interesting to notice that during a very short period of time, the simulation continues and this regime shows a loss of hyperbolicity related to the condition $K_{,X}+2X K_{,XX}=0$. This result is similar to \cite{Bernard:2019fjb}.
\item Increasing the amplitude of the initial field, the sonic horizon is present, but now the metric tends to collapse forming an event horizon with the same radius than the sonic horizon (within the accuracy of our simulation). This behavior is observed when $\beta$ parameter is not too strong. We do not observe any loss of hyperbolicity.
\item Increasing the amplitude of the scalar field, the evolution ends because of the formation of a sonic horizon, as in the first situation, but this time the dynamics of the metric seems clearly to indicate that black hole would form in the future and it seems that hyperbolicity will not be lost, the function $K_{,X}+2X K_{,XX}$ is far from vanishing.
\item For strong values of $\phi_0$ both metrics collapse at the same radius (within the accuracy of the numerical evolution). This behavior was observed for all values of $\beta$. The larger the value of $\beta$, the larger the value of $\phi_0$ producing this behavior. In this case, a BH forms and hyperbolicity is maintained until the end of the simulation.
\end{itemize}

In general, the formation of a sonic horizon indicates the formation of an event horizon in the future except in the first regime described earlier where the hyperbolicity is lost and therefore the BH never forms.

\begin{figure}
\begin{tabular}{ccccc}
\multicolumn{2}{c}{Family A} & & \multicolumn{2}{c}{Family B} \\
\includegraphics[scale=0.21]{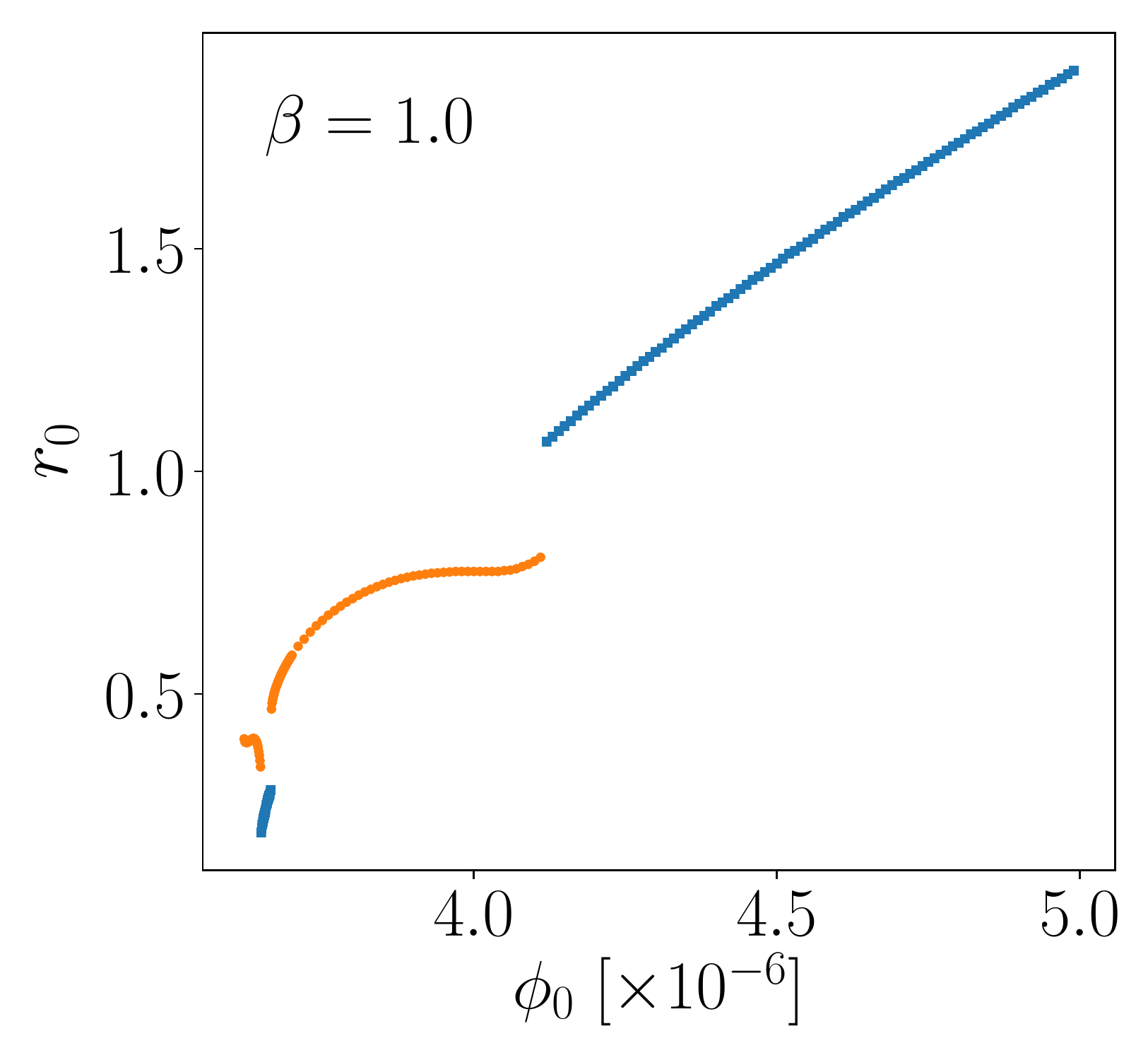}	&
\includegraphics[scale=0.21]{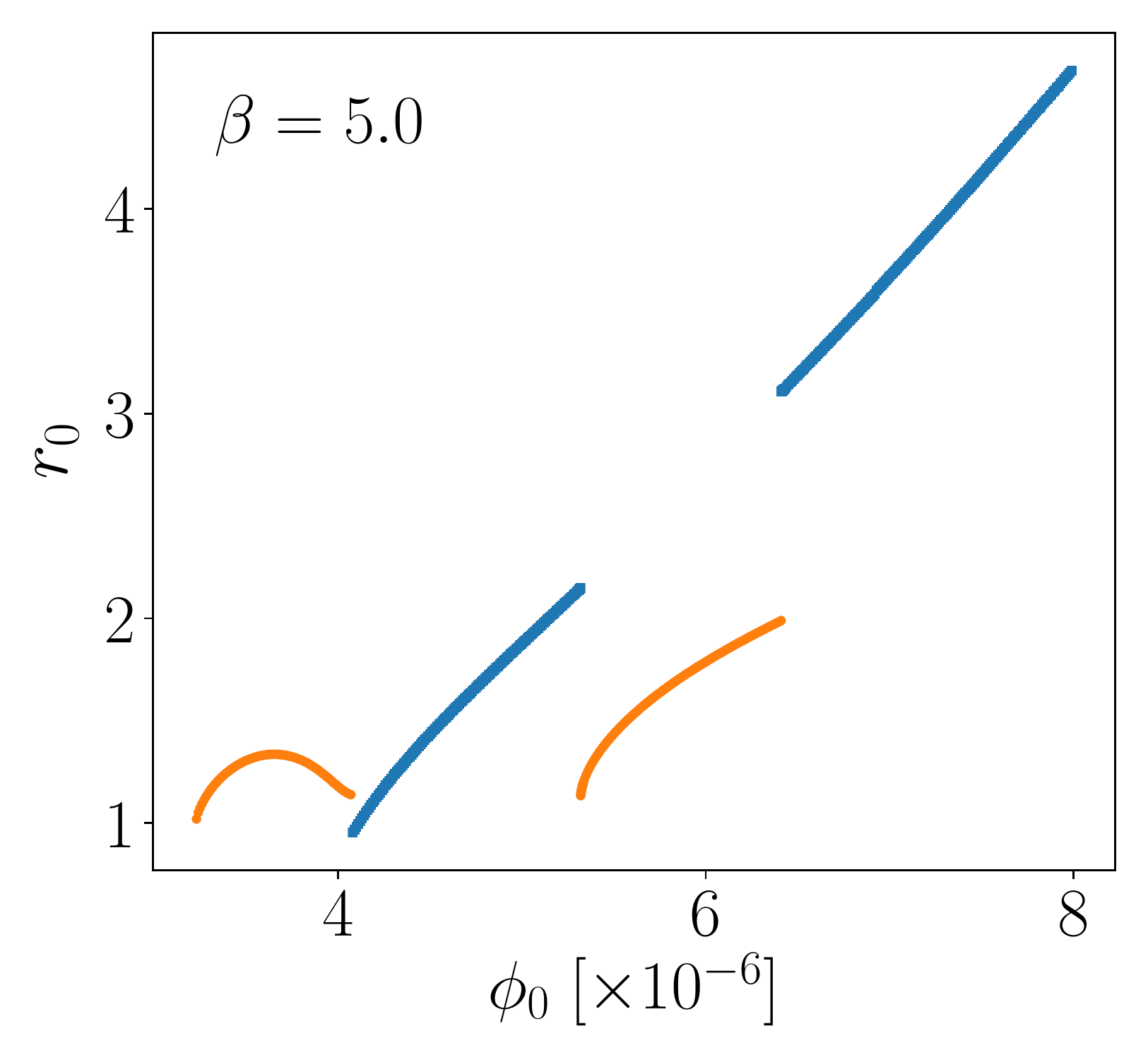}	& &	
\includegraphics[scale=0.21]{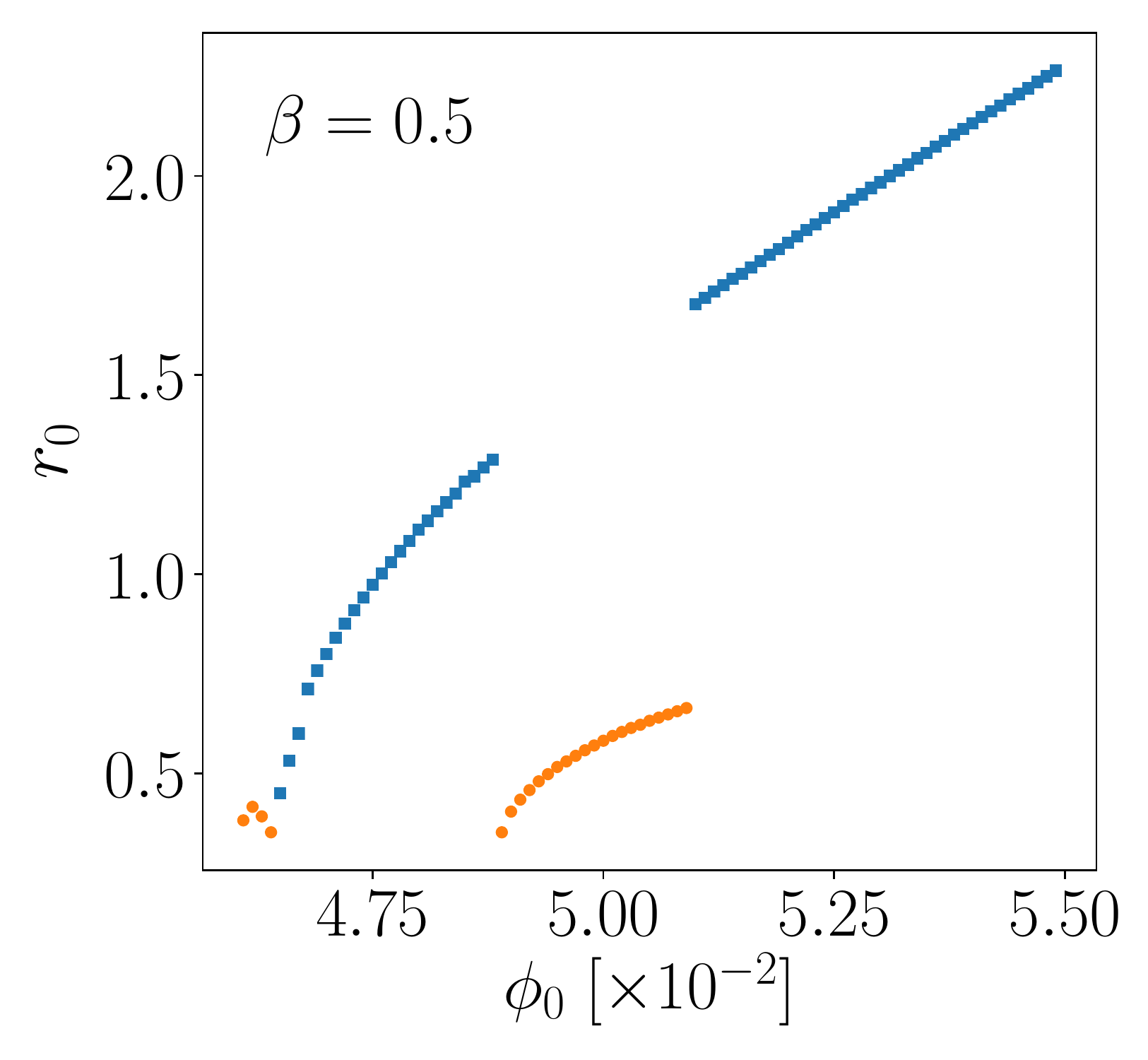}	&
\includegraphics[scale=0.21]{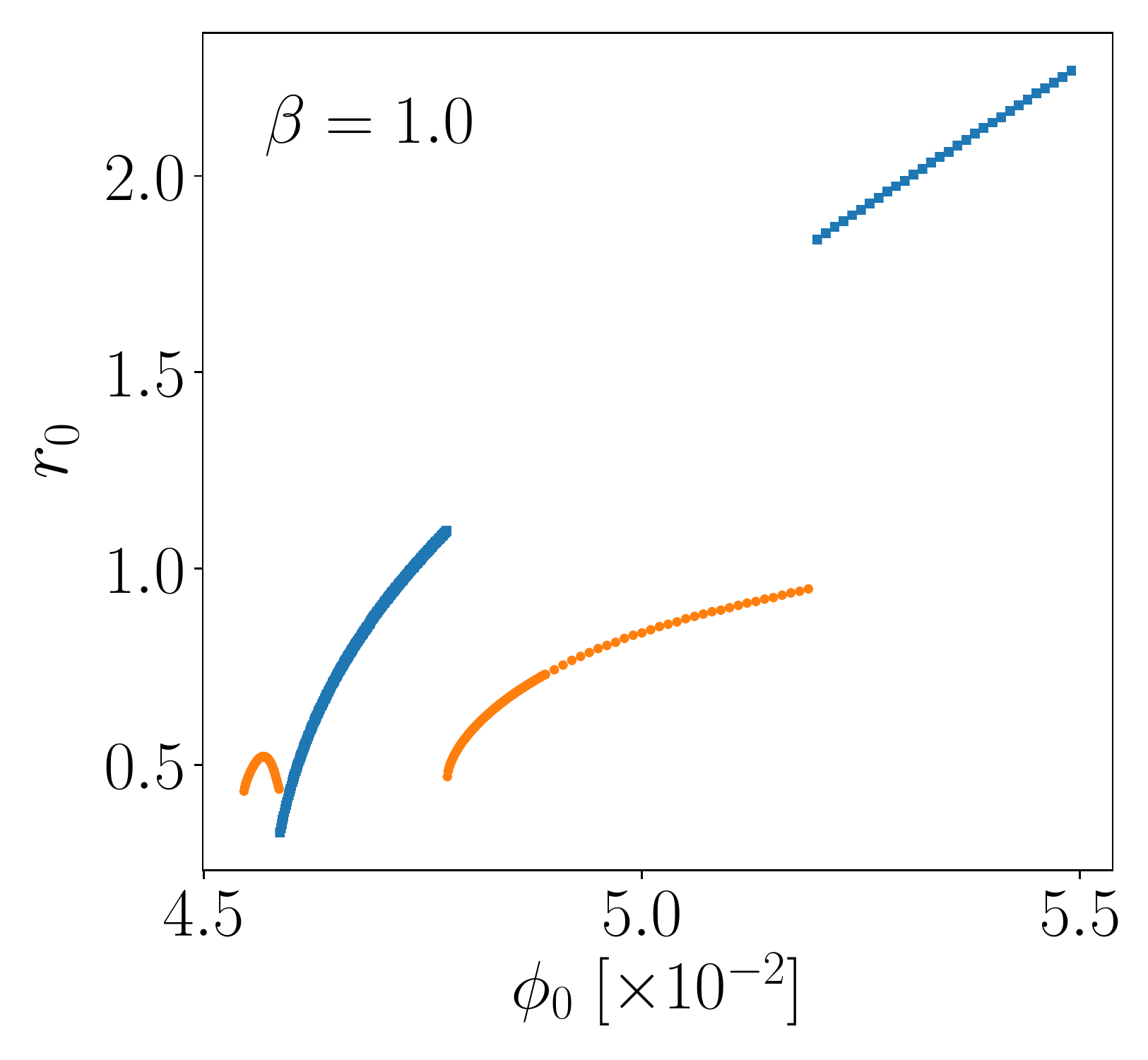}	\\	
\includegraphics[scale=0.21]{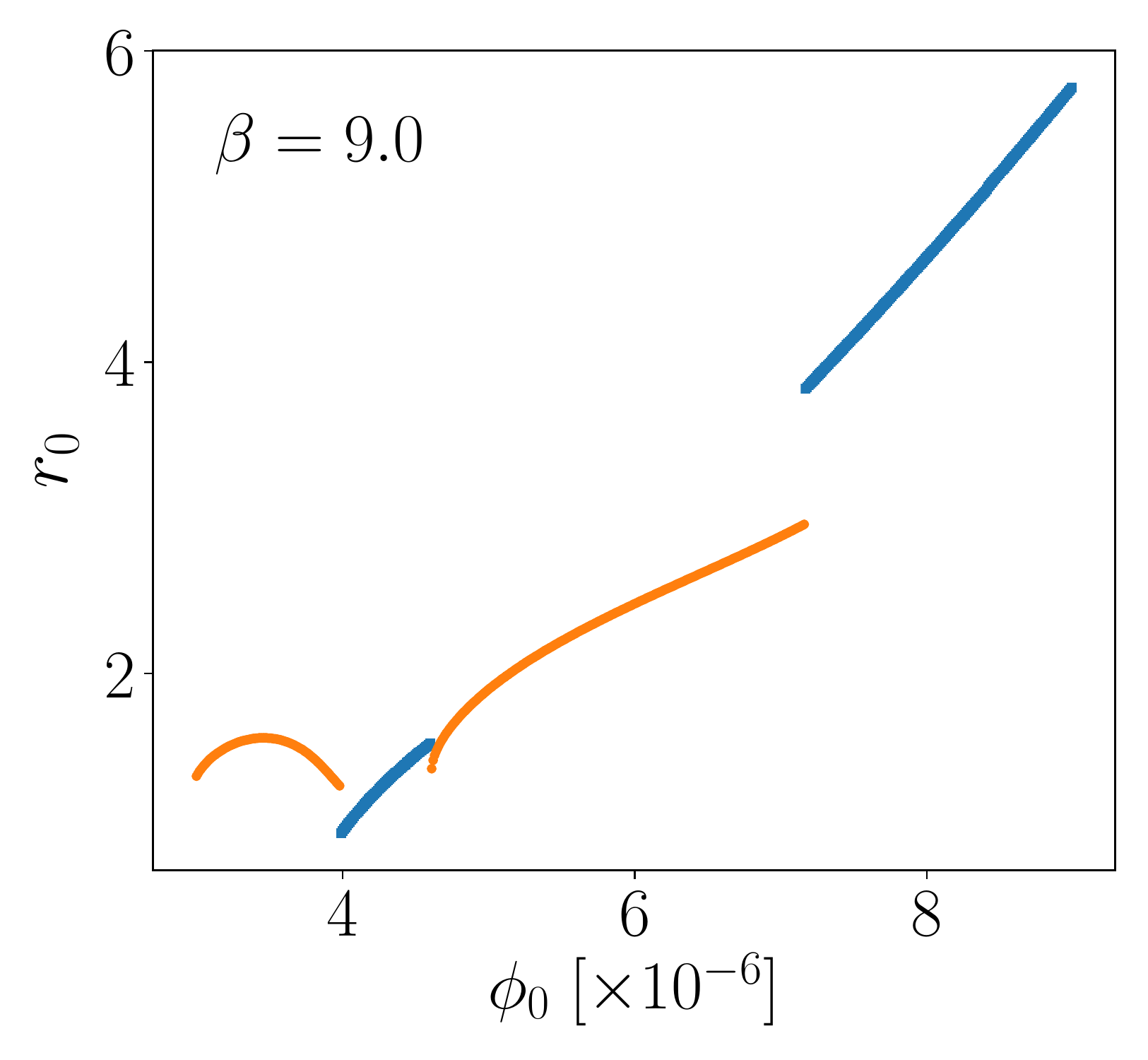}	&
\includegraphics[scale=0.21]{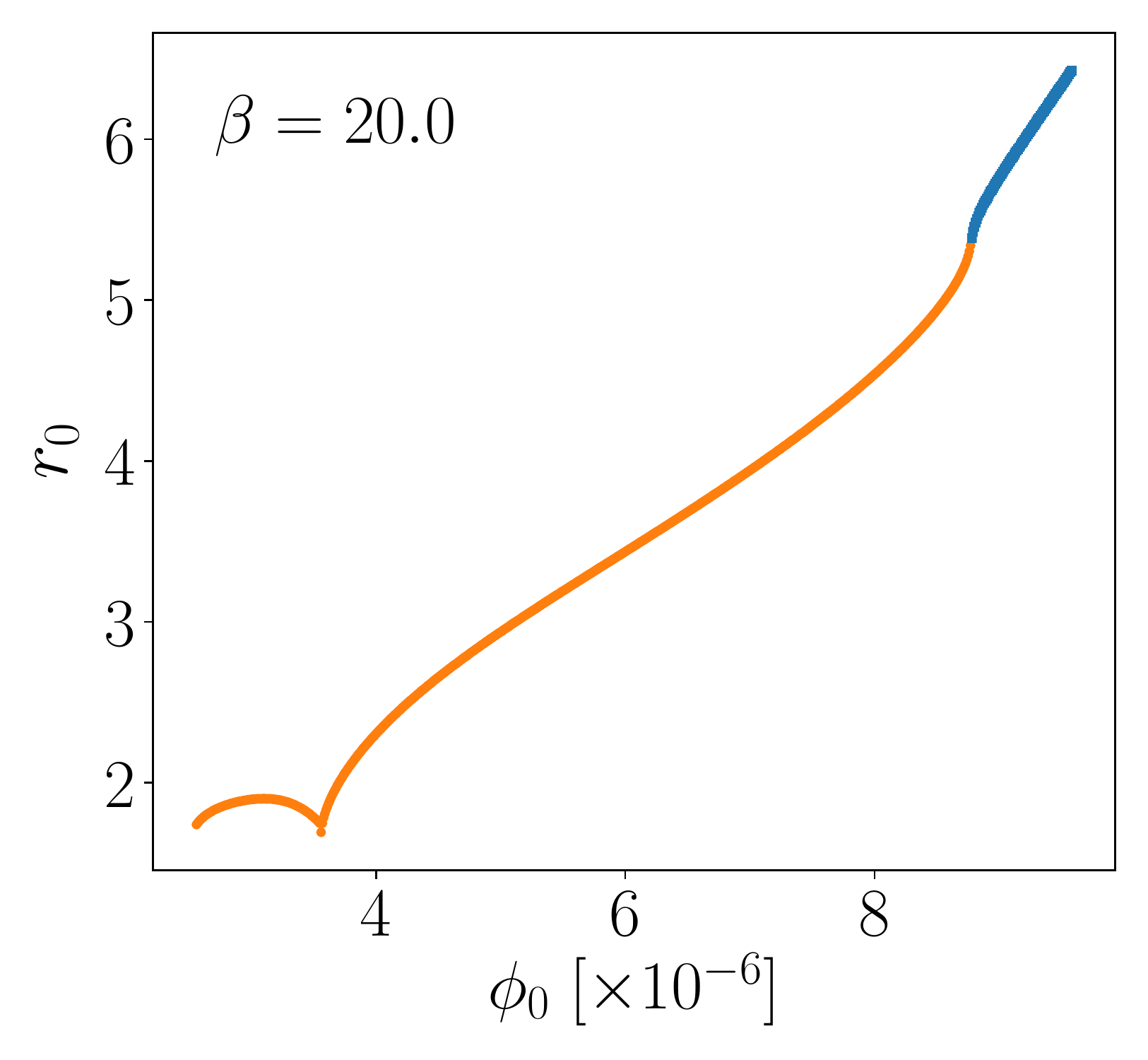}	& &	
\includegraphics[scale=0.21]{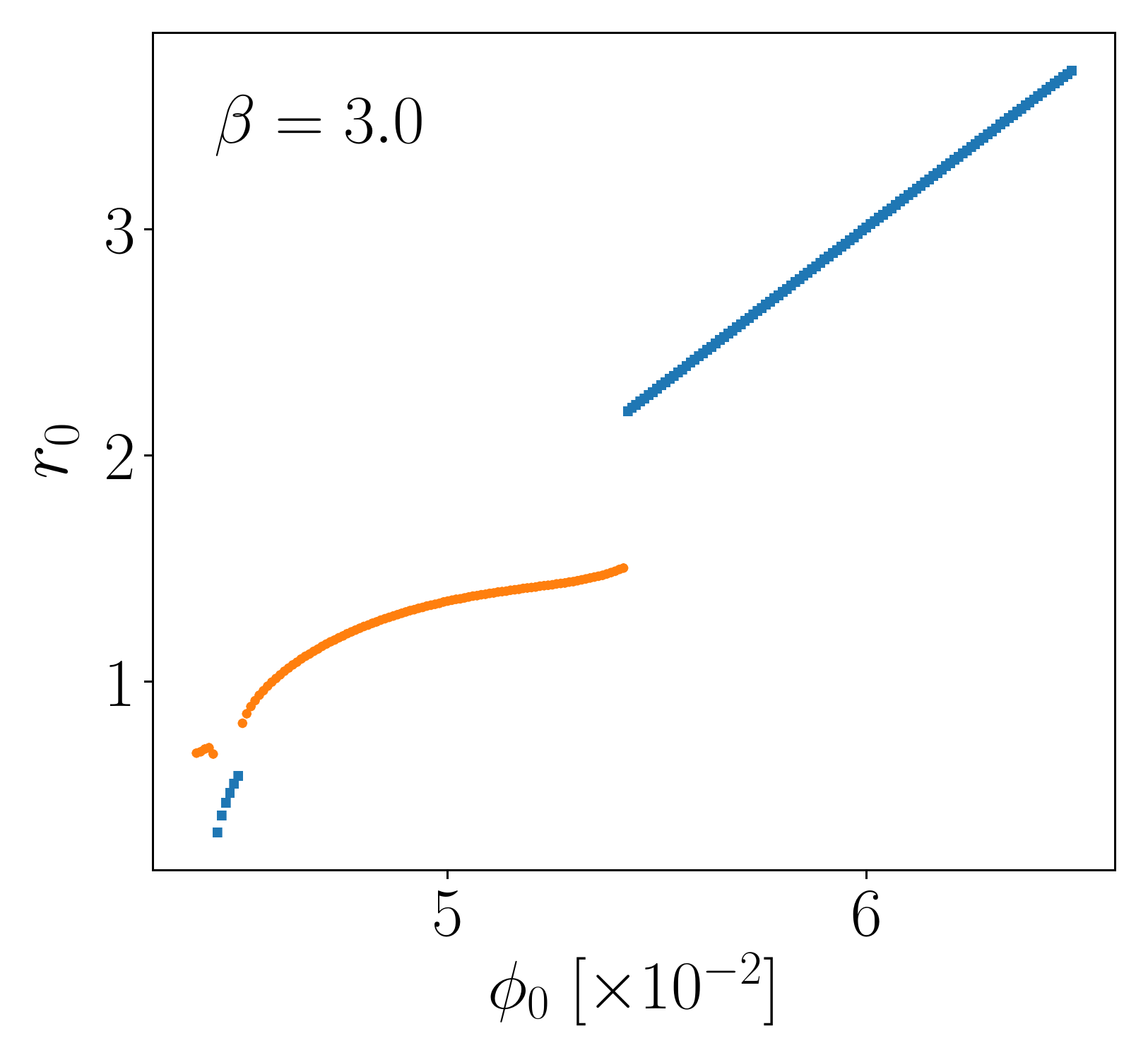}	&
\includegraphics[scale=0.21]{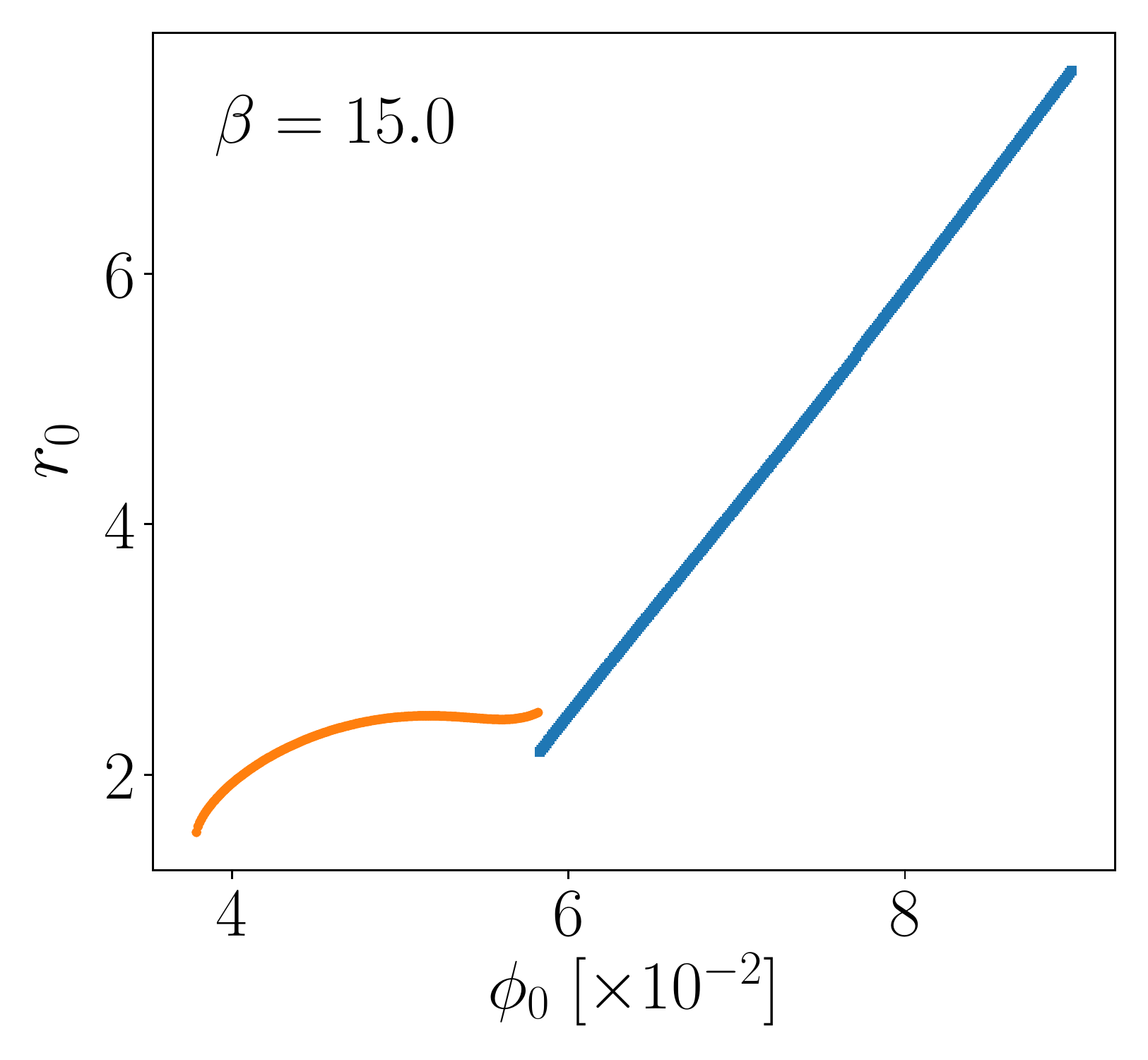}
\end{tabular}
\caption{Radius of the first horizon formed as a function of the amplitude of the initial scalar field, $\phi_0$,  for various values of $\beta$. Blue-branch represents formation of both horizons at the same time while orange-branch describes formation of sonic horizon only. Results are shown for Gaussian family (first two columns) and the Family B initial conditions (last two columns). 
\label{FamilyAB}}
\end{figure}

As shown in Fig.(\ref{Fig:4regimes}) depending on the initial value of the amplitude of the scalar field, we have either the formation of sonic horizon without the existence yet of an event horizon or both horizons (sonic and luminal) form and are indistinguishable. To illustrate these behaviors in a better way, in Fig.(\ref{FamilyAB}) we show the variation of the apparent radius defined either by the sonic horizon or by both horizons when formed simultaneously. We see e.g. for $\beta=5$ and for the Family A of initial conditions that we have 4 different regimes corresponding to the cases described in Fig.(\ref{Fig:4regimes}). The first and the third regime (in orange) corresponds to the formation of the sonic horizon while the second and the last (in blue) corresponds to the simultaneous formation of both horizons. Notice that for larger values of $\beta$ some regimes disappear.
In case of the Family A and $\beta=5$, the blue lines seem to form only 1 line if extended. In fact, we expect, in this case, the third regime where only a sonic horizon forms to evolve in time with an increasing sonic horizon until it forms the link between the second regime and the last. This behavior can be observed in all cases like e.g. $\beta=1$ for Family B, where the 2 blue lines seem clearly to be extended to each other. 

Therefore, we expect the sonic horizon to be dynamical and evolve in time until the formation of the event horizon if hyperbolicity is not lost. Because in all our simulations, the spacetime seems to converge to the Schwarzschild solution, we expect both horizons to join in the future (see Appendix \ref{Appendix}). Therefore we presume that for larger time evolution, the radius of the black hole formed should describe a continuous function of the initial condition $\phi_0$ except if hyperbolicity is lost as shown in the first branch. Surprisingly, the first branch which does not represent a BH because hyperbolicity is lost, and therefore an event horizon will never form, shows a universal behavior as if it represents the threshold of black hole formation.

\begin{figure}
\includegraphics[scale=0.45]{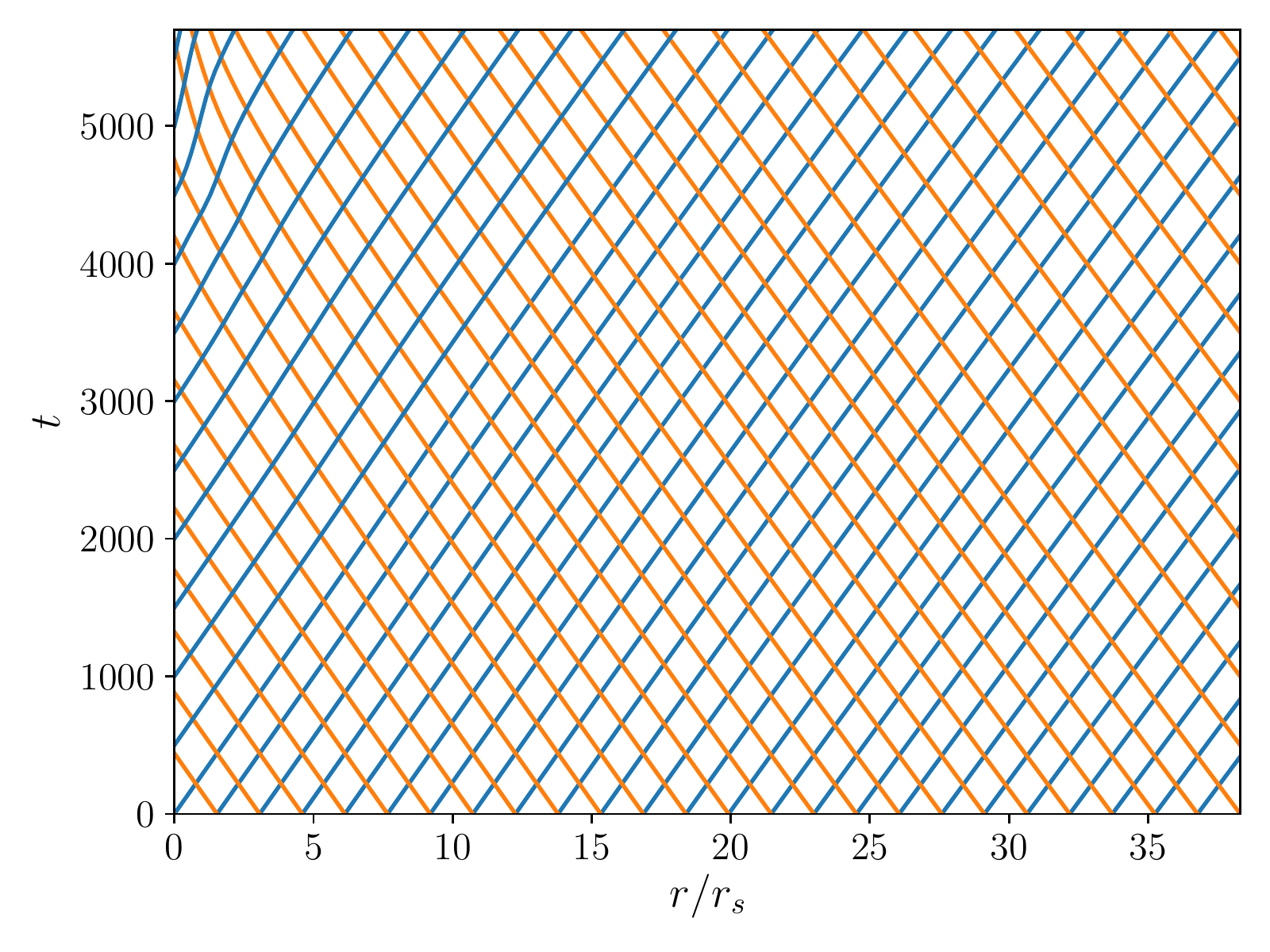}	
\includegraphics[scale=0.45]{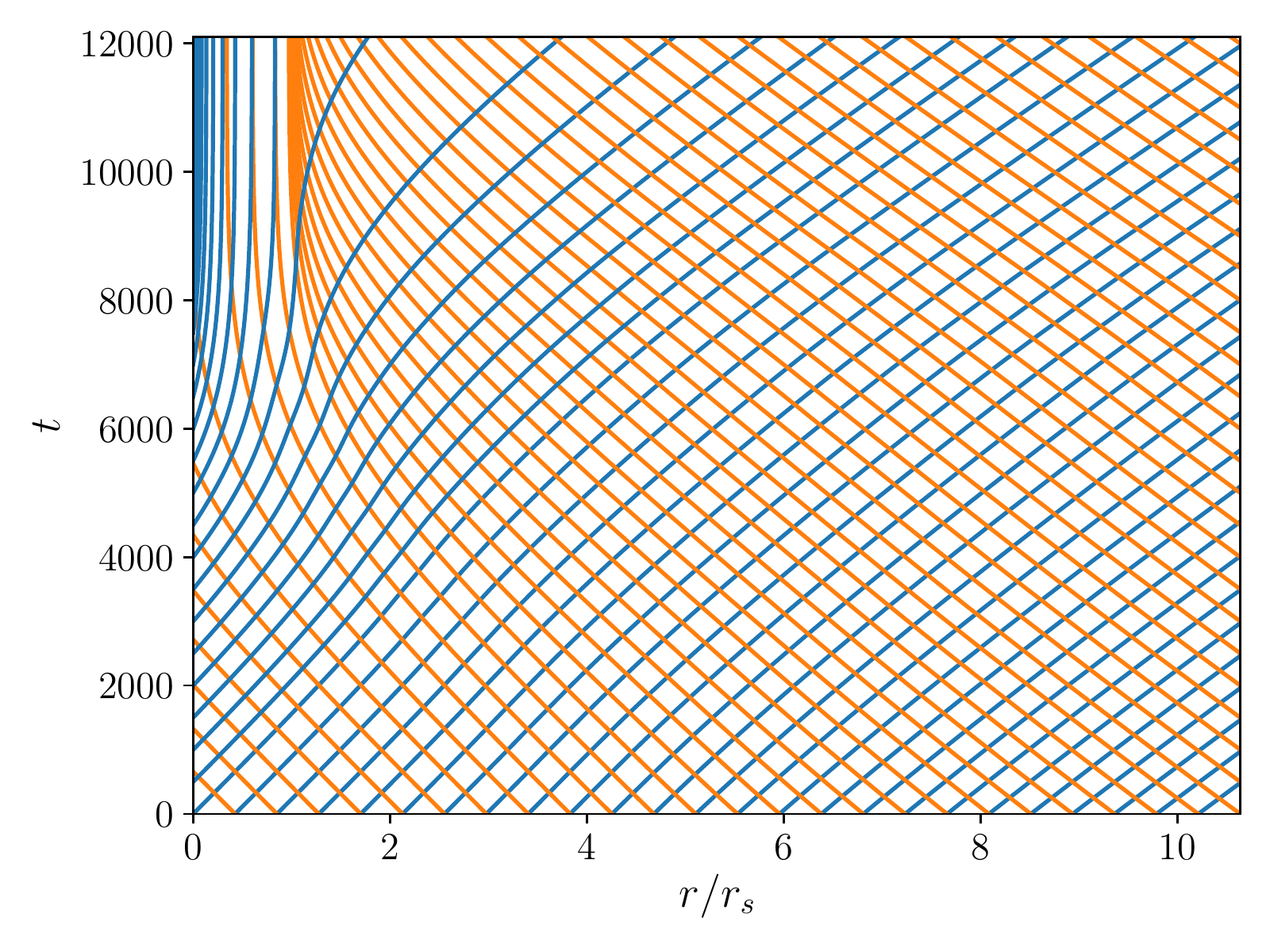}				
\caption{Integral curves of the ingoing (orange) and outgoing (blue) characteristic speed $c_\pm$ defined by eq.(\ref{char}) for $\beta=5$ and gaussian initial conditions. In the left panel, $\phi_0=3.5~10^{-6}$ and therefore the solution belongs to the first branch, where we have formation of only the sonic horizon. In the right panel, $\phi_0=8~10^{-6}$ and therefore it belongs to the last branch where a black hole forms.
\label{Fig:caustic}}
\end{figure}
We can see from Fig.(\ref{Fig:caustic}) the evolution of the characteristic line defined from eq.(\ref{char}) in $(t,r)$ coordinates. In the first case, a sonic horizon is formed but we do not expect a black hole to form as explained previously, while in the second case, an event horizon forms at normalized radius $r=1$. Recently, it was shown that in flat spacetime, these models could produce caustics \cite{Babichev:2016hys}. But as it can be seen from Fig.(\ref{Fig:caustic}), the characteristic lines do not intersect. We haven't found any formation of caustics in our simulations. In the second case, where the event horizon forms, indicated by the lines converging to $r=r_s$ of the characteristic lines, the BH is Schwarzschild.

\begin{figure}
\begin{tabular}{cc}
\includegraphics[scale=0.47]{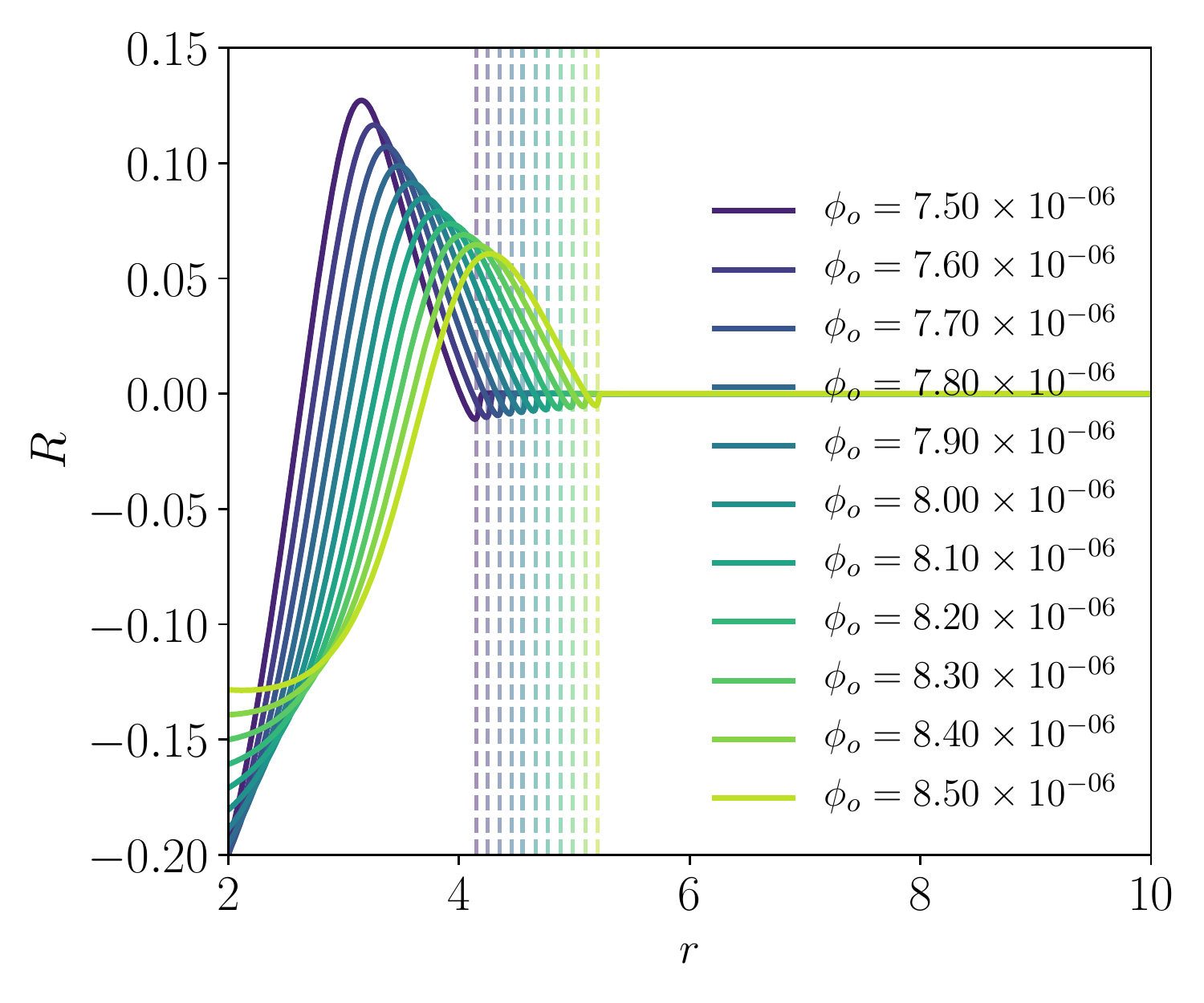} 	&
\includegraphics[scale=0.47]{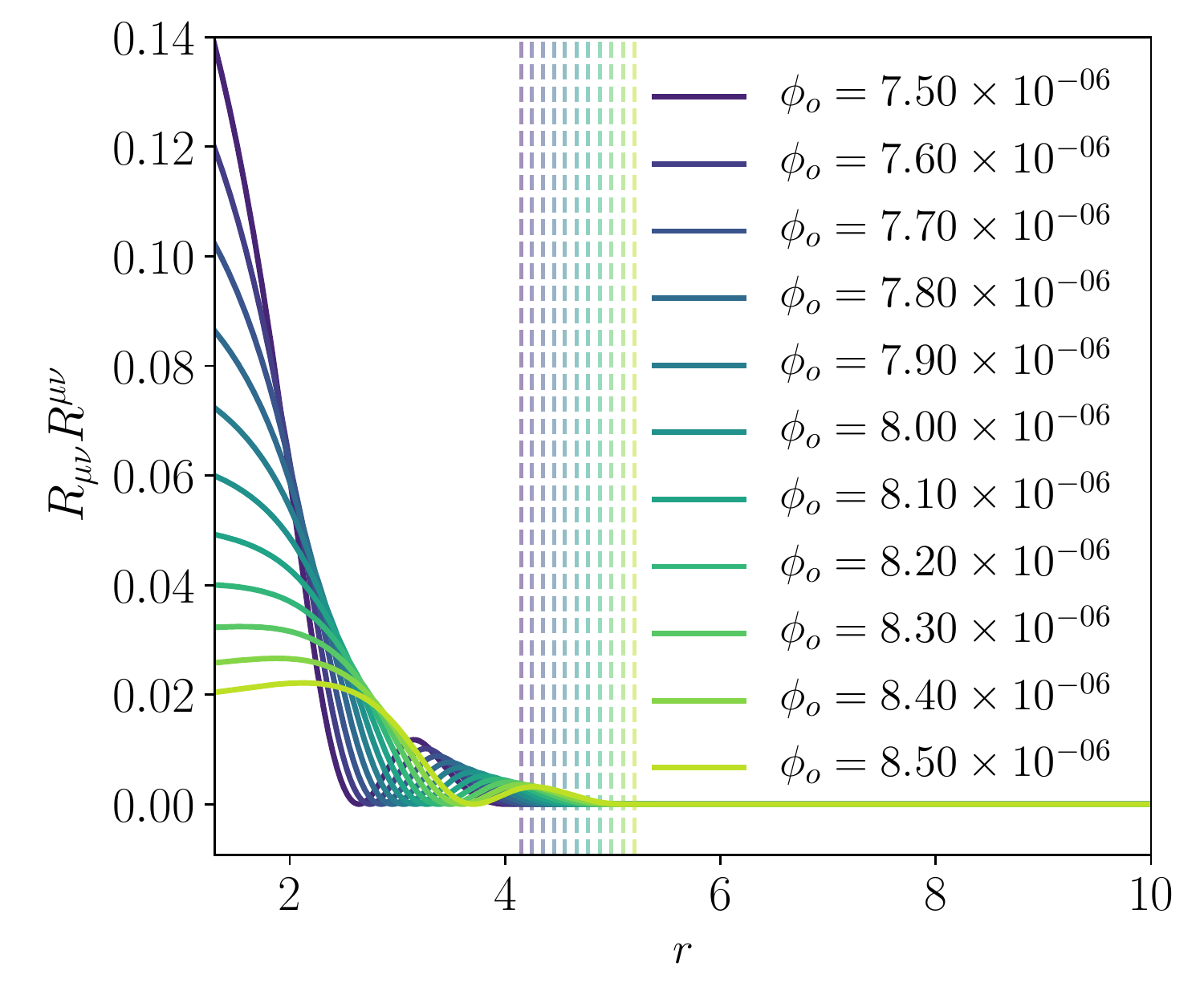}
\end{tabular}
\caption{Curvature scalar $R$ and Ricci tensor squared $R_{\mu\nu}R^{\mu\nu}$ as a function of the radial radius $r$ at last time $t$ before formation of the event horizon for $\beta=10$ and $\phi_0$ corresponding to the last branch. The vertical line represents the position of the event horizon for each initial condition $\phi_0$.}
\label{FIG:R}
\end{figure}

In fact, every time a black hole forms, the exterior solution is Schwarzschild as we can see in Fig.(\ref{FIG:R}). Considering e.g. the Family A of initial conditions
and for various values of $\phi_0$ which all correspond to the fourth branch where the event horizon forms at the same time than the sonic horizon (blue-branch in Fig.(\ref{FamilyAB})), we have represented the curvature scalar $R$ and the Ricci tensor squared $R_{\mu\nu}R^{\mu\nu}$ during the last moment of evolution before black hole formation. We see that for all values of $\phi_0$ the curvature scalar and the Ricci tensor squared vanish for $r$ larger than the event horizon indicating the formation of the Schwarzschild solution which has also been checked directly from the metric. This behavior has been observed for both families of initial conditions and for all values of $\beta$. The end state of the evolution when the event horizon is formed is the Schwarzschild spacetime. Notice also that the event horizon increases with increasing $\phi_0$ as expected.

\begin{figure}
\begin{tabular}{cc}
\includegraphics[scale=0.47]{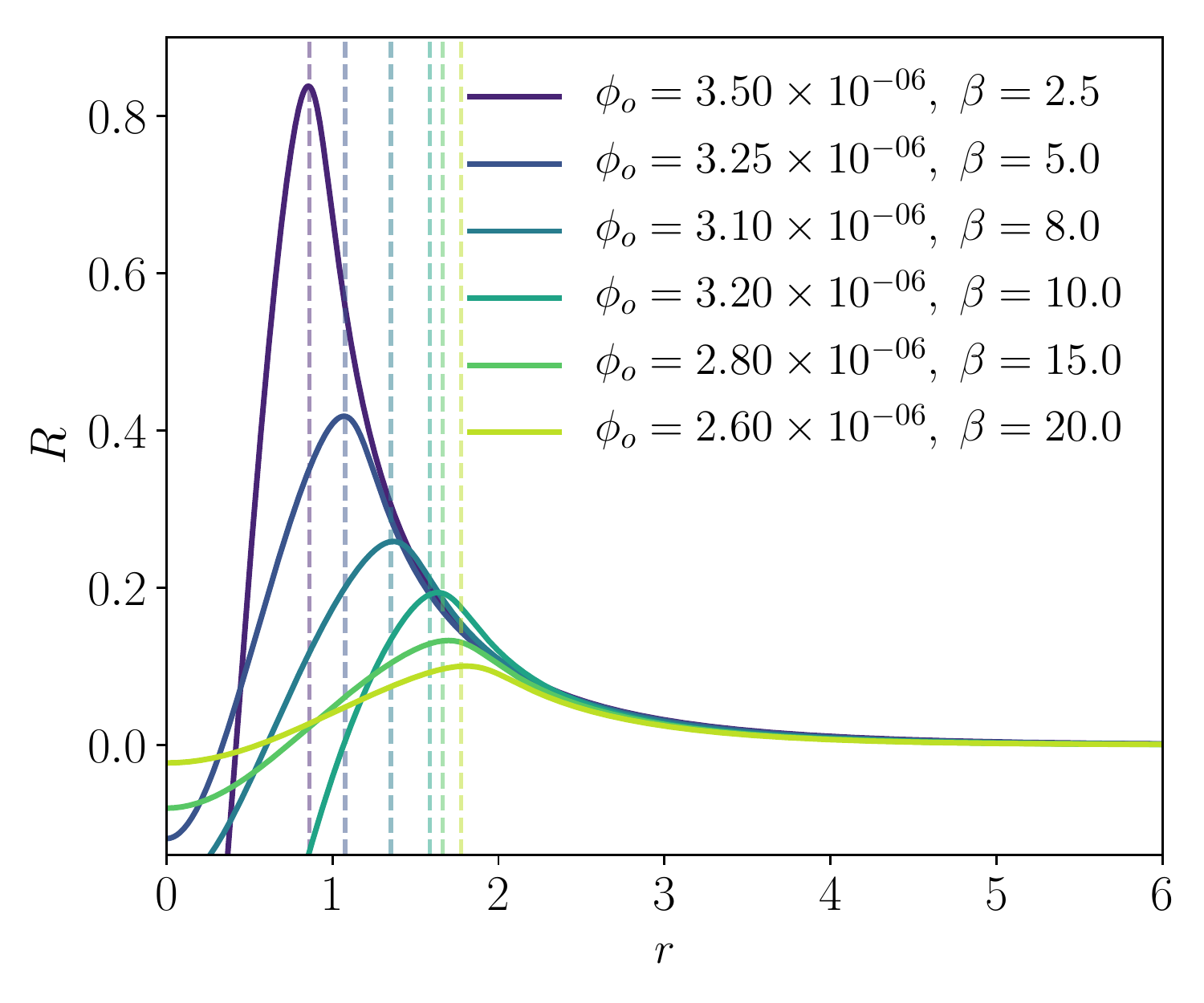} 	&
\includegraphics[scale=0.47]{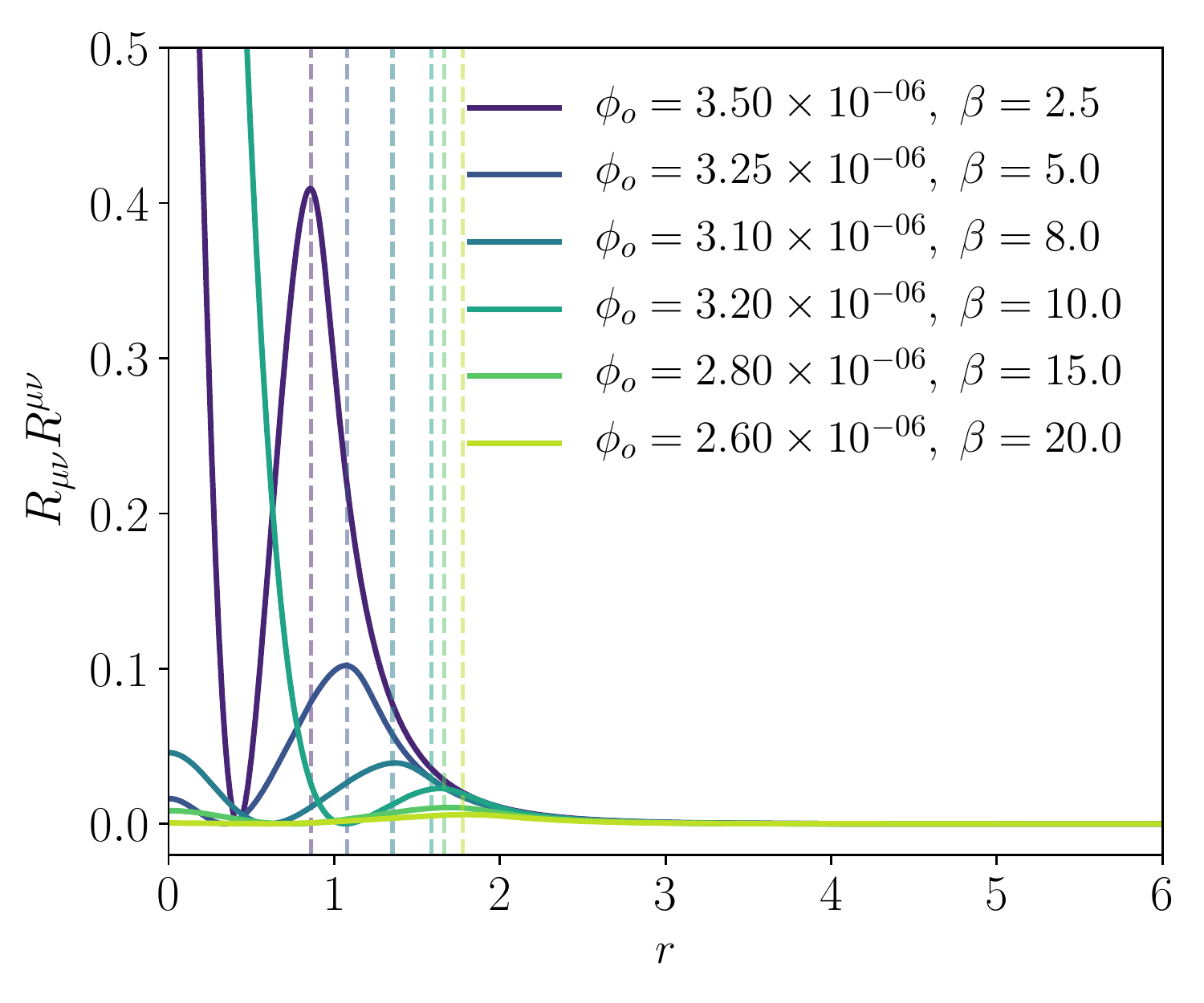}
\end{tabular}
\caption{Curvature scalar $R$ and Ricci tensor squared $R_{\mu\nu}R^{\mu\nu}$ as a function of the radial radius $r$ at last moment $t$ before formation of the sonic horizon for various values of $\beta$ and $\phi_0$ corresponding to the first branch. The vertical line represents the position of the sonic horizon.}
\label{FIG:R2}
\end{figure}

\begin{figure}
\begin{center}
\includegraphics[scale=0.8]{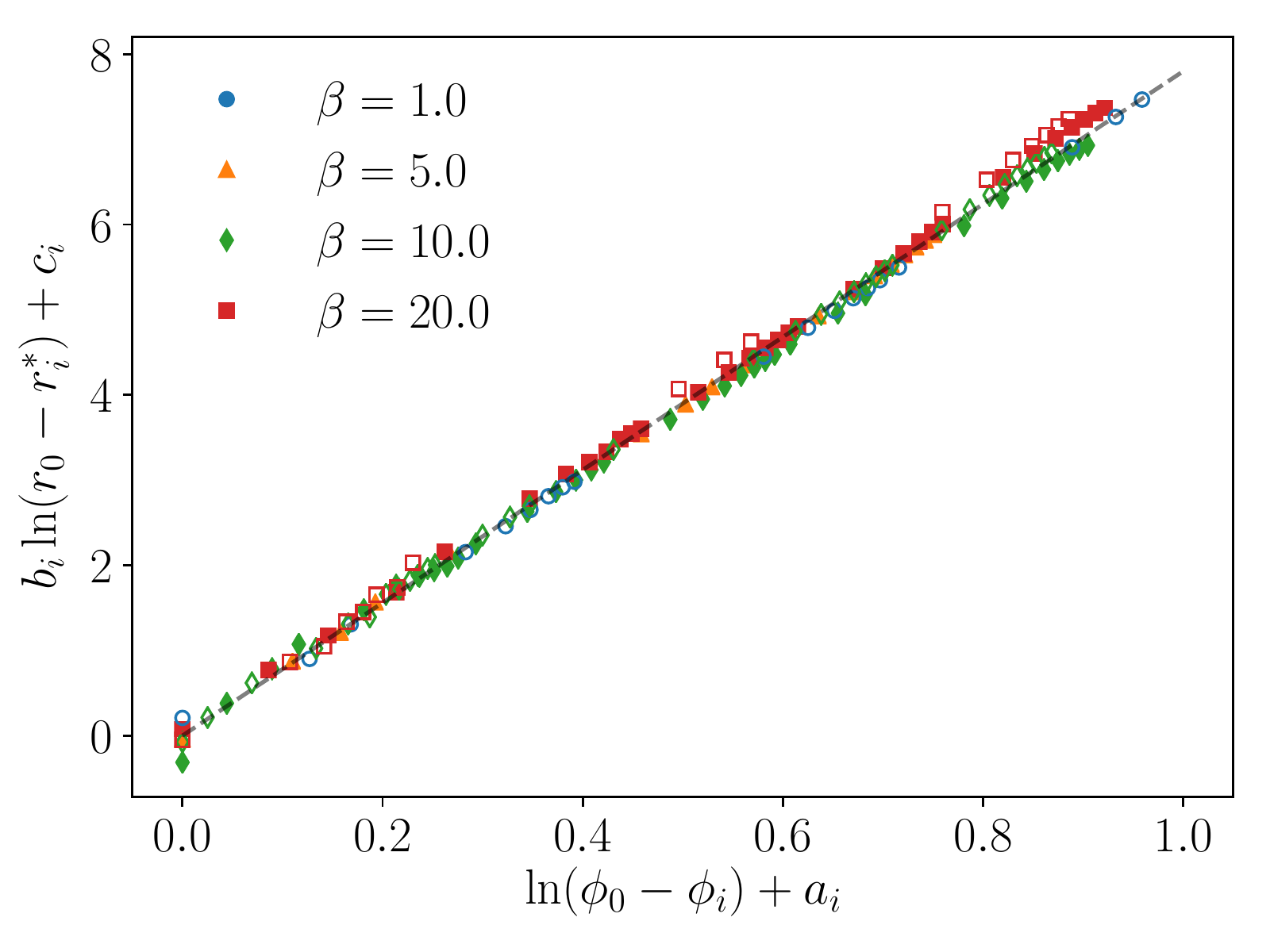}
\caption{Radius scaling relation of the sonic horizon for the two families of initial conditions and for various values of $\beta$. Data with filled markers correspond to Family A. Data with empty markers correspond to Family B.}
\label{TendencyLine}
\end{center}
\end{figure}

On the other hand, when the sonic horizon forms first, the metric is not Schwarzschild as seen in Fig.(\ref{FIG:R2}). In this case, we expect the system to continue to evolve until the formation of the Schwarzschild spacetime or a loss of hyperbolicity. This behavior should be checked with coordinates such as Gullstrand-Painlev\'e.
For a given family of initial conditions and for a given $\beta$, values of $\phi_0$ lower than the first branch produce dispersion and therefore flat spacetime while values taken within this branch produce a sonic horizon and later a loss of hyperbolicity. It is interesting to see from the curvature in Fig.(\ref{FIG:R2}) that we are still far from the Schwarzschild solution and therefore the formation of the event horizon, but very surprisingly, considering the sonic horizon we found a universal behavior. For any family of initial conditions and for any $\beta$, there exist a critical value of $\phi_0$ named $\phi_i$ in Fig.(\ref{TendencyLine}) around which the radius of the sonic horizon follows a universal behavior given by
\begin{align}
r=r_0+(\phi_0-\phi_i)^{\gamma}~,~~~\gamma\simeq 0.51
\end{align}
Because of the existence of an additional scale in our system ($\beta$), we have a non vanishing minimum radius of the black hole corresponding therefore to Type I critical phenomena \cite{Gundlach:2007gc}.

\section{Conclusions}

In this paper, we have studied gravitational collapse in K-essence models with additional shift symmetry. We have presented the various constraints for a well-defined problem at classical level and for its quantum completion which reduces to $K_{,X}>0$, $K_{,XX}>0$ and $K_{,X}+2XK_{,XX}>0$. For these theories, we generically have the formation of two horizons, an event horizon and a sonic horizon which define a limit for the propagation of the perturbations of the scalar field. For numerical purposes, we focused on a particular model defined by $K=X+\beta X^2$. We found that in the weak field regime, the scalar field disperses and spacetime is flat while in the strong field regime, we have the formation of a horizon. Two situations occur, either only a sonic horizon forms or both horizons form at the same time. In this last case, the exterior solution is always Schwarzschild and we never observed formation of caustics. In the cases where the sonic horizon formed first, we could have either a dynamics of the field showing the possible formation of a BH in the future or a loss of hyperbolicity of the equations. Very surprisingly, the lowest regime, corresponding to a situation were the field does not disperse to infinity, and which corresponds to the formation of a sonic horizon, reveals a universal behavior even if the hyperbolicity is rapidly lost after the formation of the sonic horizon. We found that in the critical limit of the formation of the sonic horizon $r_S$, a universal power-law scaling of $r_S$ appears with a critical exponent of order $0.51$ for any parameter $\beta \neq 0$. This result seems to indicate that the universal behavior is also encoded in the sonic horizon.

\appendix
\section[Stealth Scalar Field in K-essence]{Stealth Scalar Field in K-essence\footnote{We thank again the anonymous referee which suggested to look to this problem.}}
\label{Appendix}

In this appendix, we show that a non-trivial scalar field could live over a Schwarzschild background, which therefore could produce a sonic horizon at a location different from the event horizon. This type of dressing a black hole is known as stealth scalar field \cite{AyonBeato:2004ig}. Since the original paper, various solutions have been discovered, see e.g. \cite{AyonBeato:2005tu,Faraoni:2010mj,Babichev:2013cya} for a study in other theories.

When the light horizon forms and the spacetime solution is Schwarzschild, the Einstein tensor vanishes which implies a vanishing of the energy-momentum tensor. The right-hand side of equations (\ref{eq:constraint1},\ref{eq:constraint2}) is zero, from which we obtain the conditions of existence of this solution
\begin{align}
&K=0\\
&K_{,X}=0~~~ ||~~~ \Phi=\Pi=0
\end{align}
Because, $\Phi=\Pi=0$ would provide a trivial scalar field, we conclude that a necessary condition of existence of a dressed Schwarzschild black hole is $K=0$ and $K_{,X}=0$.

In our case, we studied the model $K=X+\beta X^2$ which can't comply with these conditions at the same time. Therefore, we conclude that every time the background is Schwarzschild, the scalar field is trivial, which implies that the effective metric is $\tilde{g}^{\mu\nu}=g^{\mu\nu}$, and therefore the sonic horizon coincides with the event horizon.

But for more generic models, we could have non-trivial solutions. Considering these 2 necessary conditions, $K=K_{,X}=0$, a solution would be of the form $X = X_0$ constant. 

In order to study also the regularity of the field across the event horizon, we introduce the ingoing Eddington-Finkelstein coordinate, $v=t+r^*$, where the tortoise coordinate is defined as $r^*=r+r_s \ln (\frac{r}{r_s}-1)$. In these coordinates, the equation $X=X_0$ reads as
\begin{align}
-\phi_{,v}\phi_{,r}-\frac{\alpha^2}{2}\phi_{,r}^2=X_0\,,\quad \alpha^2=1-\frac{r_s}{r}
\end{align}
whose solution is
\begin{align}
\phi=q v+F(r)
\end{align}
where
\begin{align}
F'(r)=\frac{-q \pm \sqrt{q^2-2X_0 \alpha^2}}{\alpha^2}
\end{align}
We see that for the square root to be always real, we need to impose $X_0\geq 0$ and $q^2-2X_0>0$, while for $X_0<0$ there will be always a range of $r$, near the singularity, where the function is not defined. Also in order to impose a regularity of the $F'(r)$ at the event horizon, we choose the solution
\begin{align}
F'(r)=\frac{-q + \sqrt{q^2-2X_0 \alpha^2}}{\alpha^2}
\end{align}
which near the horizon behaves as $-X_0/q$ (we chose $q>0$). Integrating the equation, we find the final solution 
\begin{align}
\label{eq:Stealth}
\phi(v,r)&=q(v-r)+\psi(r)-q r_s \ln\Bigl[q^2 r +(r_s-r) X_0 + q \psi(r)\Bigr]\nonumber\\
&+\frac{r_s(q^2-X_0)}{\sqrt{q^2-2X_0}}\ln\Bigl[r (q^2 - X_0)+X_0 (r_s-r) + \psi(r) \sqrt{q^2 - 2 X_0}\Bigr]
\end{align}
with $\psi(r)=r\sqrt{q^2-2X_0+2X_0 r_s/r}$. The scalar field is regular at the horizon but diverges at infinity even if its space derivative is finite.\\
Notice that for flat spacetime, $r_s=0$, we find from \ref{eq:Stealth}
\begin{align}
\phi(t,r)&=qt+r\sqrt{q^2-2X_0}
\end{align}
which corresponds to a wave of velocity $v=\sqrt{1-2X_0/q^2}$, as found in \cite{Vikman:2007sj}. We see also that $X_0=0$, reduces the field to $\phi=qv$.

Considering this solution, \ref{eq:Stealth}, which dresses the Schwarzschild spacetime, we can obtain the effective metric
\begin{align}
\tilde{g}^{11}=-K_{XX}(X_0)~\Bigl[q^2-2X_0+2X_0 \frac{r_s}{r}\Bigr]
\end{align}
Following the regularity conditions we imposed, $X_0>0,~q^2-2X_0>0$, the solution will not form a sonic horizon. The only sonic horizon which can be formed are for solutions which are not defined until $r=0$ ($X_0<0$). In this case, the sonic horizon is located at 
\begin{align}
r=\frac{-2X_0 r_s}{q^2-2X_0}
\end{align}
which corresponds to $\psi(r)=0$.

Notice that the effective metric is singular because $\tilde{g}^{22}=\tilde{g}^{33}=0$. To regularize this solution, we would need an angular dependence of the scalar field. But in all cases, it was claimed in \cite{deRham:2019gha} that the solution would be infinitely strongly coupled and therefore could not be trusted within the regime of validity of this effective field theory. It would be interesting to see if the recent extension \cite{Babichev:2018twg} of these theories suffer from the same problem.

\newpage

\acknowledgments
We would like to thank constructive comments by Alexander Vikman, Laura Bernard and the anonymous referee. R.G. is supported by FONDECYT project No 1171384 and Y.R.B. is supported by CONICYT Chile, scholarship No. 21180423.


\begin{thebibliography}{99}

\bibitem{Will:2014kxa} 
  C.~M.~Will,
  ``The Confrontation between General Relativity and Experiment,''
  Living Rev.\ Rel.\  {\bf 17}, 4 (2014)
  [arXiv:1403.7377 [gr-qc]].

\bibitem{Hawking:1973uf}
  S.~W.~Hawking and G.~F.~R.~Ellis,
  ``The Large Scale Structure of Space-Time,''

\bibitem{Vafa:2005ui}
  C.~Vafa,
  hep-th/0509212.

\bibitem{Palti:2019pca}
  E.~Palti,
  Fortsch.\ Phys.\  {\bf 67} (2019) no.6,  1900037
  doi:10.1002/prop.201900037
  [arXiv:1903.06239 [hep-th]].

\bibitem{Agrawal:2018own}
  P.~Agrawal, G.~Obied, P.~J.~Steinhardt and C.~Vafa,
  Phys.\ Lett.\ B {\bf 784} (2018) 271
  doi:10.1016/j.physletb.2018.07.040
  [arXiv:1806.09718 [hep-th]].
  
\bibitem{Giesler:2019uxc}
  M.~Giesler, M.~Isi, M.~A.~Scheel and S.~Teukolsky,
  Phys.\ Rev.\ X {\bf 9} (2019) no.4,  041060
  doi:10.1103/PhysRevX.9.041060
  [arXiv:1903.08284 [gr-qc]].

\bibitem{Garousi:2000tr}
  M.~R.~Garousi,
  Nucl.\ Phys.\ B {\bf 584} (2000) 284
  doi:10.1016/S0550-3213(00)00361-8
  [hep-th/0003122].

\bibitem{Bergshoeff:2000dq}
  E.~A.~Bergshoeff, M.~de Roo, T.~C.~de Wit, E.~Eyras and S.~Panda,
  JHEP {\bf 0005} (2000) 009
  doi:10.1088/1126-6708/2000/05/009
  [hep-th/0003221].

\bibitem{Sen:2002in}
  A.~Sen,
  JHEP {\bf 0207} (2002) 065
  doi:10.1088/1126-6708/2002/07/065
  [hep-th/0203265].

\bibitem{Born:1934gh}
  M.~Born and L.~Infeld,
  Proc.\ Roy.\ Soc.\ Lond.\ A {\bf 144} (1934) no.852,  425.
  doi:10.1098/rspa.1934.0059

\bibitem{ArmendarizPicon:1999rj} 
  C.~Armendariz-Picon, T.~Damour and V.~F.~Mukhanov,
  Phys.\ Lett.\ B {\bf 458}, 209 (1999)
  doi:10.1016/S0370-2693(99)00603-6
  [hep-th/9904075].

\bibitem{ArmendarizPicon:2000ah} 
  C.~Armendariz-Picon, V.~F.~Mukhanov and P.~J.~Steinhardt,
  Phys.\ Rev.\ D {\bf 63}, 103510 (2001)
  doi:10.1103/PhysRevD.63.103510
  [astro-ph/0006373].

\bibitem{ArmendarizPicon:2005nz} 
  C.~Armendariz-Picon and E.~A.~Lim,
  JCAP {\bf 0508}, 007 (2005)
  doi:10.1088/1475-7516/2005/08/007
  [astro-ph/0505207].

\bibitem{Choptuik:1992jv} 
  M.~W.~Choptuik,
  Phys.\ Rev.\ Lett.\  {\bf 70}, 9 (1993).
  doi:10.1103/PhysRevLett.70.9

\bibitem{May:1966zz} 
  M.~M.~May and R.~H.~White,
  Phys.\ Rev.\  {\bf 141}, 1232 (1966).
  doi:10.1103/PhysRev.141.1232

\bibitem{Healy:2013xia}
  J.~Healy and P.~Laguna,
  Gen.\ Rel.\ Grav.\  {\bf 46} (2014) 1722
  doi:10.1007/s10714-014-1722-2
  [arXiv:1310.1955 [gr-qc]].
 
\bibitem{Brady:1997fj} 
  P.~R.~Brady, C.~M.~Chambers and S.~M.~C.~V.~Goncalves,
  Phys.\ Rev.\ D {\bf 56}, R6057 (1997)
  doi:10.1103/PhysRevD.56.R6057
  [gr-qc/9709014].

\bibitem{Hawley:2000dt} 
  S.~H.~Hawley and M.~W.~Choptuik,
  Phys.\ Rev.\ D {\bf 62}, 104024 (2000)
  doi:10.1103/PhysRevD.62.104024
  [gr-qc/0007039].

\bibitem{Evans:1994pj} 
  C.~R.~Evans and J.~S.~Coleman,
  Phys.\ Rev.\ Lett.\  {\bf 72}, 1782 (1994)
  doi:10.1103/PhysRevLett.72.1782
  [gr-qc/9402041].
 
\bibitem{Sorkin:2005vz} 
  E.~Sorkin and Y.~Oren,
  Phys.\ Rev.\ D {\bf 71}, 124005 (2005)
  doi:10.1103/PhysRevD.71.124005
  [hep-th/0502034].

\bibitem{Nicolis:2008in} 
  A.~Nicolis, R.~Rattazzi and E.~Trincherini,
  Phys.\ Rev.\ D {\bf 79}, 064036 (2009)
  doi:10.1103/PhysRevD.79.064036
  [arXiv:0811.2197 [hep-th]].
  
\bibitem{Deffayet:2009wt} 
  C.~Deffayet, G.~Esposito-Farese and A.~Vikman,
  Phys.\ Rev.\ D {\bf 79}, 084003 (2009)
  doi:10.1103/PhysRevD.79.084003
  [arXiv:0901.1314 [hep-th]].

\bibitem{Kobayashi:2011nu} 
  T.~Kobayashi, M.~Yamaguchi and J.~Yokoyama,
  Prog.\ Theor.\ Phys.\  {\bf 126}, 511 (2011)
  doi:10.1143/PTP.126.511
  [arXiv:1105.5723 [hep-th]].
  
\bibitem{Horndeski:1974wa} 
  G.~W.~Horndeski,
  Int.\ J.\ Theor.\ Phys.\  {\bf 10}, 363 (1974).
  doi:10.1007/BF01807638

\bibitem{Zumalacarregui:2013pma} 
  M.~Zumalacárregui and J.~García-Bellido,
  Phys.\ Rev.\ D {\bf 89}, 064046 (2014)
  doi:10.1103/PhysRevD.89.064046
  [arXiv:1308.4685 [gr-qc]].
  
\bibitem{Gleyzes:2014dya} 
  J.~Gleyzes, D.~Langlois, F.~Piazza and F.~Vernizzi,
  Phys.\ Rev.\ Lett.\  {\bf 114}, no. 21, 211101 (2015)
  doi:10.1103/PhysRevLett.114.211101
  [arXiv:1404.6495 [hep-th]].
  
\bibitem{Gleyzes:2014qga} 
  J.~Gleyzes, D.~Langlois, F.~Piazza and F.~Vernizzi,
  JCAP {\bf 1502}, 018 (2015)
  doi:10.1088/1475-7516/2015/02/018
  [arXiv:1408.1952 [astro-ph.CO]].
  
\bibitem{Papallo:2017qvl} 
  G.~Papallo and H.~S.~Reall,
  Phys.\ Rev.\ D {\bf 96}, no. 4, 044019 (2017)
  doi:10.1103/PhysRevD.96.044019
  [arXiv:1705.04370 [gr-qc]].

\bibitem{Akhoury:2008nn} 
  R.~Akhoury, C.~S.~Gauthier and A.~Vikman,
  JHEP {\bf 0903}, 082 (2009)
  doi:10.1088/1126-6708/2009/03/082
  [arXiv:0811.1620 [astro-ph]].

\bibitem{Babichev:2018twg}
E.~Babichev, S.~Ramazanov and A.~Vikman,
JCAP \textbf{11}, 023 (2018)
doi:10.1088/1475-7516/2018/11/023
[arXiv:1807.10281 [gr-qc]].

\bibitem{Garriga:1999vw} 
  J.~Garriga and V.~F.~Mukhanov,
  Phys.\ Lett.\ B {\bf 458}, 219 (1999)
  doi:10.1016/S0370-2693(99)00602-4
  [hep-th/9904176].

\bibitem{Wald:1984rg} 
  R.~M.~Wald,
  ``General Relativity,''
  doi:10.7208/chicago/9780226870373.001.0001

\bibitem{ArkaniHamed:2003uy} 
  N.~Arkani-Hamed, H.~C.~Cheng, M.~A.~Luty and S.~Mukohyama,
  JHEP {\bf 0405}, 074 (2004)
  doi:10.1088/1126-6708/2004/05/074
  [hep-th/0312099].

\bibitem{Babichev:2018uiw} 
  E.~Babichev, C.~Charmousis, G.~Esposito-Farèse and A.~Lehébel,
  Phys.\ Rev.\ D {\bf 98}, no. 10, 104050 (2018)
  doi:10.1103/PhysRevD.98.104050
  [arXiv:1803.11444 [gr-qc]].

\bibitem{Adams:2006sv} 
  A.~Adams, N.~Arkani-Hamed, S.~Dubovsky, A.~Nicolis and R.~Rattazzi,
  JHEP {\bf 0610}, 014 (2006)
  doi:10.1088/1126-6708/2006/10/014
  [hep-th/0602178].

\bibitem{Nicolis:2009qm} 
  A.~Nicolis, R.~Rattazzi and E.~Trincherini,
  JHEP {\bf 1005}, 095 (2010)
  Erratum: [JHEP {\bf 1111}, 128 (2011)]
  doi:10.1007/JHEP05(2010)095, 10.1007/JHEP11(2011)128
  [arXiv:0912.4258 [hep-th]].

\bibitem{Melville:2019wyy} 
  S.~Melville and J.~Noller,
  Phys.\ Rev.\ D {\bf 101}, no. 2, 021502 (2020)
  doi:10.1103/PhysRevD.101.021502
  [arXiv:1904.05874 [astro-ph.CO]].

\bibitem{Leonard:2011ce} 
  C.~D.~Leonard, J.~Ziprick, G.~Kunstatter and R.~B.~Mann,
  JHEP {\bf 1110}, 028 (2011)
  doi:10.1007/JHEP10(2011)028
  [arXiv:1106.2054 [gr-qc]].

\bibitem{Babichev:2006vx}
E.~Babichev, V.~F.~Mukhanov and A.~Vikman,
JHEP \textbf{09}, 061 (2006)
doi:10.1088/1126-6708/2006/09/061
[arXiv:hep-th/0604075 [hep-th]].

\bibitem{Hawking:1973qla} 
  S.~W.~Hawking,
  ``The event horizon'' (1973),
  Les Houches Summer School of Theoretical Physics : Black Holes


\bibitem{Ashtekar:2004cn} 
  A.~Ashtekar and B.~Krishnan,
  Living Rev.\ Rel.\  {\bf 7}, 10 (2004)
  doi:10.12942/lrr-2004-10
  [gr-qc/0407042].
    
\bibitem{Courant:1962}
R.~Courant and D.~Hilbert, 
Methods of Mathematical Physics V. II (1962).    
    
\bibitem{Ripley:2019hxt} 
  J.~L.~Ripley and F.~Pretorius,
  Phys.\ Rev.\ D {\bf 99}, no. 8, 084014 (2019)
  doi:10.1103/PhysRevD.99.084014
  [arXiv:1902.01468 [gr-qc]].

\bibitem{Rendall:2005fv}
A.~D.~Rendall,
Class. Quant. Grav. \textbf{23}, 1557-1570 (2006)
doi:10.1088/0264-9381/23/5/008
[arXiv:gr-qc/0511158 [gr-qc]].

\bibitem{Bernard:2019fjb} 
  L.~Bernard, L.~Lehner and R.~Luna,
  Phys.\ Rev.\ D {\bf 100}, no. 2, 024011 (2019)
  doi:10.1103/PhysRevD.100.024011
  [arXiv:1904.12866 [gr-qc]].

\bibitem{Gundlach:2007gc}
  C.~Gundlach and J.~M.~Martin-Garcia,
  Living Rev.\ Rel.\  {\bf 10} (2007) 5
  doi:10.12942/lrr-2007-5
  [arXiv:0711.4620 [gr-qc]].

\bibitem{AyonBeato:2004ig}
E.~Ayon-Beato, C.~Martinez and J.~Zanelli,
Gen. Rel. Grav. \textbf{38}, 145-152 (2006)
doi:10.1007/s10714-005-0213-x
[arXiv:hep-th/0403228 [hep-th]].
%
\bibitem{AyonBeato:2005tu}
E.~Ayon-Beato, C.~Martinez, R.~Troncoso and J.~Zanelli,
Phys. Rev. D \textbf{71}, 104037 (2005)
doi:10.1103/PhysRevD.71.104037
[arXiv:hep-th/0505086 [hep-th]].

\bibitem{Faraoni:2010mj}
V.~Faraoni and A.~F.~Moreno,
Phys. Rev. D \textbf{81}, 124050 (2010)
doi:10.1103/PhysRevD.81.124050
[arXiv:1006.1936 [gr-qc]].

\bibitem{Babichev:2013cya}
E.~Babichev and C.~Charmousis,
JHEP \textbf{08}, 106 (2014)
doi:10.1007/JHEP08(2014)106
[arXiv:1312.3204 [gr-qc]].
  
\bibitem{Vikman:2007sj}
A.~Vikman,
``K-essence: cosmology, causality and emergent geometry,''

\bibitem{Babichev:2016hys}
E.~Babichev,
JHEP \textbf{04}, 129 (2016)
doi:10.1007/JHEP04(2016)129
[arXiv:1602.00735 [hep-th]].

\bibitem{deRham:2019gha}
C.~de Rham and J.~Zhang,
Phys. Rev. D \textbf{100}, no.12, 124023 (2019)
doi:10.1103/PhysRevD.100.124023
[arXiv:1907.00699 [hep-th]].

\end{thebibliography}
\end{document}